\documentclass[aps,prx,twocolumn,longbibliography,superscriptaddress]{revtex4-2}
\usepackage{graphicx}
\usepackage{latexsym}
\usepackage[nointegrals]{wasysym}
\usepackage{amssymb}
\usepackage{amsfonts}
\usepackage{upgreek}
\usepackage{bm}
\usepackage{multirow}
\usepackage{mathtools}
\usepackage{nicematrix}
\usepackage{enumitem}
\usepackage{color}
\usepackage{hyperref}
\usepackage{physics}
\usepackage{bm}
\usepackage{youngtab}
\usepackage{ytableau}
\usepackage{tcolorbox}
\usepackage{pifont}

\hypersetup{
colorlinks = true,
linkcolor = [rgb]{0.70,0.13,0.13},
citecolor = [rgb]{0.13,0.55,0.13},
urlcolor = [rgb]{0.25, 0.41, 0.88}}

\newcommand{\bI}{\mathbb{I}}
\newcommand{\cmark}{{\color{blue} \ding{51}}}%
\newcommand{\xmark}{{\color{red} \ding{55}}}%

\newcommand{\U}{\mathrm{U}}
\newcommand{\SU}{\mathrm{SU}}
\newcommand{\Z}{\mathrm{Z}}

\newcommand{\br}{{\bm{r}}}
\newcommand{\bsigma}{{\bm{\sigma}}}

\newcommand{\bK}{{\bm{K}}}

\newcommand{\bq}{{\bm{q}}}
\newcommand{\bn}{{\bm{n}}}

\newcommand{\bG}{{\bm{G}}}

\newcommand{\bk}{{\bm{k}}}
\newcommand{\cC}{{\cal{C}}}

\newcommand{\cT}{\mathcal{T}}

\newcommand{\cK}{\mathcal{K}}
\newcommand{\TK}{\,\textrm{K}}
\newcommand{\KM}{\textbf{K}}
\newcommand{\meV}{\,\textrm{meV}}

\newcommand{\secref}[1]{Sec.\,\ref{#1}}
\newcommand{\appref}[1]{Appendix\,\ref{#1}}
\newcommand{\eqnref}[1]{Eq.\,\eqref{#1}}
\newcommand{\figref}[1]{Fig.\,\ref{#1}}
\newcommand{\tabref}[1]{Tab.\,\ref{#1}}

\newcommand*{\PRX}[1]{\textcolor{black}{{#1}}}

\begin{document}

\title{Fermionic Monte Carlo study of a realistic model of twisted bilayer graphene}

\author{Johannes S. Hofmann}
\affiliation{Department of Condensed Matter Physics, Weizmann Institute of Science, Rehovot 76100, Israel}

\author{Eslam Khalaf}
\affiliation{Department of Physics, Harvard University, Cambridge, MA 02138, USA}

\author{Ashvin Vishwanath}
\affiliation{Department of Physics, Harvard University, Cambridge, MA 02138, USA}

\author{Erez Berg}
\affiliation{Department of Condensed Matter Physics, Weizmann Institute of Science, Rehovot 76100, Israel}

\author{Jong Yeon Lee}\thanks{jongyeon@kitp.ucsb.edu}
\affiliation{Kavli Institute for Theoretical Physics, University of California, Santa Barbara, CA 93106-4030, USA}

\date{\today}
\begin{abstract}
The rich phenomenology of twisted bilayer graphene (TBG) near the magic angle is believed to arise from electron correlations in topological flat bands. An unbiased approach to this problem is highly desirable, but also particularly challenging, given the  multiple electron flavors, the topological obstruction to defining tight-binding models and the long-ranged Coulomb interactions.  While numerical simulations of realistic models have thus far been confined to zero temperature, typically excluding some spin or valley species, analytic progress has relied on fixed point models away from the realistic limit. Here we present for the first time unbiased Monte Carlo simulations of realistic models of magic angle TBG at charge-neutrality.  We establish the absence of a sign problem for this model in a  momentum space approach, and describe a computationally tractable formulation that applies even on breaking chiral symmetry and including band dispersion. Our results include (i)  the emergence of an insulating  Kramers inter-valley coherent ground state  in  competition with a correlated semi-metal phase, (ii) detailed temperature evolution of order parameters and electronic spectral functions which reveal a `pseudogap' regime, in which gap features are established at a  higher temperature than the onset of order and (iii) predictions for electronic tunneling spectra and their evolution with temperature. Our results pave the way towards uncovering the physics of magic angle graphene through exact simulations of over  a hundred electrons across a wide temperature range. 
\end{abstract}
\maketitle

\section{Introduction}

The discovery of interaction-driven insulating and superconducting phases in twisted bilayer graphene (TBG) \cite{PabloMott, PabloSC} has inspired intense experimental and theoretical effort aiming to understand the nature of these phases and the relationship between them \cite{MacDonald2011, Santos, Balents18, Po2018, IsobeFu, Thomson18, YouAV, KangVafekPRX, Xie2020, Nandkishore, Kivelson, PhilipPhillips,phononMacDonald, phononLianBernevig, Zou2018, Bitan19, Zhang2018, Khalaf2019, Mora19, Cea19, fang2019angledependent, Bi2019Strain, Lee19, tbg_hubbard_simplified, BultnickKhalaf2020, kwan2021kekule, KangVafekDMRG, Soejima2020, tbg_ladder_DMRG, chatterjee2020skyrmion, Hejazi2021, Liu2019, tbgRG2020, tbgED2021, tbgHubbard2021, KhalafSC, FengWang_ARPES, Efetov_ARPES}. Such an understanding remains extremely challenging due to the strongly interacting nature of the problem combined with the non-trivial band topology \cite{Po2018, Zou2018,  StiefelWhitney, Song18, po2018faithful, Hejazi19,Song_21} which obstructs a conventional lattice description. In addition, the presence of spin and valley results in a very large manifold of possible competing states which further complicates the analysis. Finally,  remarkable phenomena are further observed at finite temperature, ranging from a flavor cascade \cite{Zondiner_2020,Wong_2020} to the Pomeranchuk effect \cite{tbg_pomeranchuk, tbg_pomeranchuk2} and unusual `strange metal' scaling of resistivity with temperature \cite{Cao_2020,Polshyn_2019} which clamor for explanation.   

To tackle this problem, various approaches have been employed which can be divided into two categories. First, there are real-space approaches based on deriving Hubbard-like models \cite{KoshinoFu, KangVafekPRX, UchoaPRL}. Such models have been analyzed in strong coupling approaches extended with perturbation theory \cite{KangVafekPRL, UchoaPRL} in addition to numerical methods such as density matrix renormalization group (DMRG) \cite{VafekMengDMRG} and QMC \cite{tbgHubbard2021}. However, all these approaches have to deal with the inherent problem that the band topology is obscured in the real space model and the symmetries are not naturally represented. Furthermore, it is not always straightforward to relate the parameters of these models to the realistic problem. 
The other class of approaches employs a momentum space continuum description based on the Bistritzer-MacDonald (BM) model~\cite{Bistritzer2011}. These approaches include strong coupling perturbative approaches \cite{BultnickKhalaf2020, SoftModes, PrincetonTBGIV}, self-consistent Hartree-Fock  \cite{Xie2020, Liu2021, BultnickKhalaf2020, Guinea2018, Guinea2019, Guinea2020, DaiHF, ZhangZhang}, DMRG \cite{KangVafekDMRG, Soejima2020,ParkerDMRG}, exact diagonalization (ED) \cite{PrincetonTBGVI, MacdonaldED} and many others \cite{IsobeFu, ChubukovTBG, ChubukovVanHove}. Each of these approaches comes with its own limitations; strong coupling perturbative approaches requires an extrapolation to  access the intermediate coupling regime which is likely to be of experimental relevance, mean-field approaches are biased towards Slater determinant states and overestimate symmetry-breaking tendencies, DMRG is largely restricted to zero temperature properties, and ED can only access very small system sizes.

\begin{figure*}[t]
    \centering
     \includegraphics[width = 1 \textwidth]{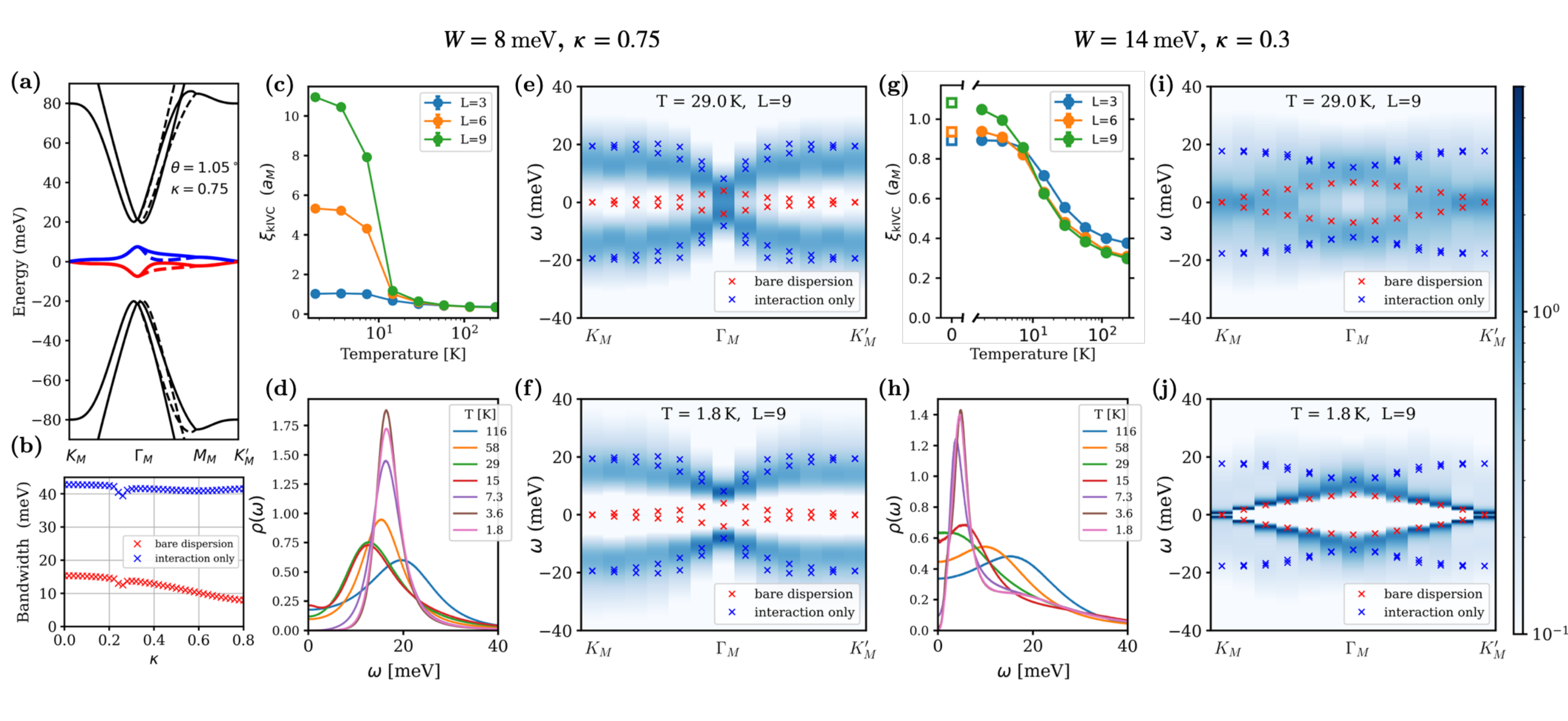}
    \caption{\label{fig:sum} 
    (a) Electron-hole excitation spectrum at charge neutrality
    [eigenvalues of $\hat{H}_0$ in Eq.~\eqref{eq:ham_QMC}]. In the DQMC simulations, we project the Hamiltonian onto the middle two narrow bands (red and blue). Solid (Dashed) lines are for + ($-$) valley. 
    (b) Bandwidth of $\hat{H}_0$ as function of $\kappa = w_0/w_1$ (red), along with the bandwidth of the excitation spectrum in the interaction-only limit, where $\hat{H}_0$ is neglected (blue).
    (c-f) DQMC results for
    $\kappa = 0.75$, for which $W=8\meV$: (c) K-IVC correlation lengths for two system sizes $L=3,6$ and $9$ in the unit of moir\'e lattice vector, (d) single-particle density of state $\rho(\omega)=\int \dd^2 \bk\, A(\omega,\bk)$, (e,f)  spectral function $A(\bm{k},\omega)$ along a cut through the moir\'e Brillouin zone for $T = 29$K and $1.8$K, respectively. The red crosses denote the single particle dispersion $\hat{H}_0$, while the blue crosses represent the single particle dispersion in the interaction only limit. 
    (g-j) Same as panels (c-f) for $\kappa = 0.3$ ($W=14\meV$). The parameters used in the simulations are given in Sec.~\ref{sec:parameters}. \PRX{In (g), the empty square represents the zero temperature data obtained by a projective variant of the QMC method~\cite{sugiyama86_proj,sorella89_proj}. 
    }}
\end{figure*}

In this work, we overcome these limitations by using a determinant quantum Monte Carlo (DQMC) method to study the behavior of the strongly interacting flat bands in TBG at charge neutrality (CN) ($\nu=0$) where we establish the absence of the sign problem.
Our approach is based on a momentum space description which enables it to accurately capture non-trivial topology of flat bands and evade the problems that plague lattice real-space descriptions. In addition, it allows us to handle the long-range Coulomb interaction and is ideally suited to address the physics at finite temperatures. Experimentally, the behavior of TBG at CN (in the absence of substrate alignment) is sample-dependent with some samples exhibiting semi-metallic transport \cite{PabloMott, Dean-Young} while others showing an activated behavior pointing to an insulating gap \cite{Efetov2019, Efetov2020}. Other probes such as scanning tunneling microscopy (STM) report a strong reconstruction of density of states and enhancement of the separation between van Hove peaks at CN, but, in the absence of further input, does not conclusively establish a gap  \cite{CaltechSTM,Andrei2019,Kerelsky2019,Yazdani2019}. The  different behavior between samples could arise from substrate alignment, disorder\cite{AliceaDisorder} or strain which was argued to induce competition between a correlated insulator and a nematic semimetal \cite{Liu2021, ParkerDMRG}. In the absence of these effects, Ref.~\cite{BultnickKhalaf2020} have proven that the ground state is necessarily an insulator in the limit of small dispersion. On the other hand, for the physically realistic parameters where interactions themselves give rise to a significant dispersion, such a correlated insulator can be destabilized. Therefore, unbiased numerical simulations are needed even to establish the phase diagram of pristine twisted bilayer graphene. 

Furthermore, access to the physics at finite temperature can shed light on a host of phenomena mentioned previously. More broadly, TBG represents a new kind of correlated insulator which occurs in topological bands, and revealing the detailed formation of such a state, as we outline here, can have  far reaching consequences.

The rest of the paper is organized as follows. In \secref{sec:sum}, we summarize the main results of the paper. 
In \secref{sec:TBG}, we examine the TBG Hamiltonian in detail and define the interacting Hamiltonian to be used in DQMC. As pointed out in \cite{Xie2020, Liu2021, BultnickKhalaf2020}, the bare kinetic energy receives corrections in the interacting problem which needs to be duly incorporated. 
In \secref{sec:Sign} we revisit the sign-problem in DQMC. We elucidate the role played by a particle-hole symmetry in sign problem. Using this, we show that the sign-problem is absent in TBG at charge neutrality under a minor approximation, 
 which amounts to neglecting the small angle rotation of the Dirac matrices \cite{Hejazi19,Song18}. 
In \secref{sec:result}, we present DQMC results with various parameter, system sizes, and temperatures. 
In \secref{sec:outlook}, we conclude the paper with an outlook.

\section{Summary of results}
\label{sec:sum}

Our starting point is the continuum model describing the moir\'e bands in TBG~\cite{Bistritzer2011} at charge neutrality, supplemented by long-range Coulomb interactions. This model respects all the relevant symmetries and the non-trivial topology of the low-energy bands. For computational simplicity, we keep only the pair of bands closest to the Fermi level [see \figref{fig:sum}(a)], and project out all the other bands, assumed to be either empty or full~\footnote{Including additional bands does not introduce a sign problem, and can in principle be incorporated in the DQMC simulations. However, the computational complexity increases as the number of bands to the third power.}. Crucially, we take into account the renormalization of the single-particle dispersion as a result of interactions~\cite{Xie2020, Liu2021, BultnickKhalaf2020}, as described in Sec.~\ref{sec:many-body}. As we shall see, this effect can significantly impact the nature of the ground state.

Our key observation is that the continuum model for TBG at charge neutrality is nearly free of the fermion sign problem in quantum Monte Carlo, due to its approximate particle-hole symmetry (PHS). Under a slight deformation of the model -- neglecting the rotation of the Pauli matrices in the individual Dirac Hamiltonians of the two graphene sheets by the twist angle $\theta$ -- this particle-hole symmetry becomes exact, and the model is completely sign-problem free. Given the smallness of the magic angle $\theta \approx 1.05^\circ$, the magnitude of the neglected term is $\sim 2\%$ of the total Hamiltonian. \PRX{We note that the PHS enhances the degeneracy of different symmetry-breaking states in the flat-band limit only, while finite dispersion lifts this degeneracy\cite{BultnickKhalaf2020}.}

The resulting PHS is represented as an anti-unitary operator acting on the many-body Hilbert space. In addition to transforming an electron into a hole, it also flips the valley, sublattice, and layer indices [see Eq.~\eqref{eq:PHS}]. This PHS, combined with a $U(1)$ spin or valley conservation symmetry, allows for a sign-problem free formulation of the DQMC method (Sec.~\ref{sec:Sign}). As a benchmark, we simulated the system in the interaction-only case where the single-particle dispersion is artificially turned off. In this case, our simulations reproduce the exact results  for the ground state energy, the nature of the ground state manifold \cite{BultnickKhalaf2020}, and the single-particle excitation spectrum.

We then turn to the realistic case, where the single-particle dispersion is included. Our main results are summarized in~\figref{fig:sum}(c--j). One of key parameters in the BM model, which changes both the bandwidth and the wavefunctions of the active bands, is the ratio between the intra-sublattice (A--A) and inter-sublattice (A--B) tunneling matrix elements between the two layers. We denote these matrix elements by $w_0$ and $w_1$, respectively. The ratio $\kappa = w_0/w_1$ is expected to be less than unity due to lattice relaxation~\cite{Koshino2017, fang2019angledependent}. We have performed DQMC simulations for two values of $\kappa$, $0.75$ and $0.3$. For $\kappa = 0.75$, we find that the ground state is a gapped Kramers inter-valley coherent (K-IVC) state~\cite{Bultinck19}, characterized by a spontaneous hybridization between the two valleys and preserving an effective time-reversal symmetry. The correlation function of the K-IVC order parameter as a function of temperature is shown in Fig.~\ref{fig:sum}(c) for three system sizes, showing a rapid growth of the correlation length below $T\sim 10\TK$.

Additional insight into the nature of the K-IVC state is gained by calculating the single-particle excitation spectrum, using the maximum entropy method~\cite{MaxEnt2004}. Fig.~\ref{fig:sum}(d) shows the single-particle density of states, $\rho(\omega)$, summed over both bands and spin/valley flavors, at different temperatures. $\rho(\omega)$ is closely related to the spectrum measured in scanning tunneling spectroscopy (STM) experiments. At the lowest temperature, $T=1.8 \TK$, the spectrum shows a clear gap of about $10\meV$ for either particle or hole excitations. Surprisingly, however, a gap-like structure in $\rho(\omega)$ is apparent already at temperatures as high as $T\approx 60\TK$, where the correlation length of the K-IVC order is around one moir\'e lattice spacing. We refer to this behavior as the formation of a `pseudogap', i.e., a spectral gap that appears with no accompanying long-range order. 

Panels (e),(f) show the spectral function $A(\bm{k},\omega)$  for $\kappa=0.75$ along a cut in the Brillouin zone at two representative temperatures, $T=1.8\TK$ and $29\TK$.  The dispersion at $1.8\TK$ is remarkably close to the spectrum of the single-particle excitation of the interaction-only model where $\hat{H}_0$ is neglected (blue crosses), for which exact results are available. The agreement between the interaction-only and DQMC dispersions is surprising, since the bare dispersion bandwidth $W$ is significant, $W\approx 8\meV$ for $\kappa=0.75$ (the single-particle dispersion used in the DQMC simulation is indicated by the red crosses). It is also worth noting that the minimum of the dispersion is at the center of the moir\'e Brillouin zone (the $\Gamma_M$ point)~\cite{BultnickKhalaf2020}, in contrast to the single-particle dispersion which has Dirac points at corners of the moir\'e Brillouin zone, $K_M$ and $K'_M$. At the higher temperature, $T=29\TK$, the peaks in the dispersion are broader~\footnote{This broadening may be an artifact of the maximum entropy algorithm, whose intrinsic frequency resolution is proportional to temperature, or may represent the physical lifetime of the excitations at finite $T$.}, but the main features -- including the gap at the $\Gamma_M$ point -- are still clearly visible, despite the absence of long-range order. Note, the spectral function can be accessed by ARPES measurements, the measurement of which has been initiated in recent experiments  \cite{FengWang_ARPES, Efetov_ARPES}.

The behavior for $\kappa=0.3$ in \figref{fig:sum}(g,h) is dramatically different from that of $\kappa=0.75$. In this case, the correlation length of the K-IVC order parameter remains short and nearly system size-independent down to the lowest temperatures. \PRX{This result is corroborated for $T=0$ using a projective variant of the QMC algorithm~\cite{sugiyama86_proj,sorella89_proj} as visualized by the open boxes in \figref{fig:sum}(g).} The system is semi-metallic with Dirac points at $K_M$ and $K'_M$, as can be seen from the spectral function, $A(\bm{k},\omega)$, at $T=1.8\TK$ (panel j). The small gap of about $0.7\meV$ at the $K_M$, $K'_M$ points is due to the finite-size charging energy; its value decreases as $1/L^2$ with the system size. The spectral function at $T=29\TK$ is much broader, although the main features of the dispersion are already visible.  

The fact that the K-IVC state is suppressed with decreasing $\kappa$ may seem surprising, since in the BM model, the bandwidth vanishes at the magic angle for $\kappa=0$ (the `chiral limit'~\cite{Tarnopolsky}). However, we find that the bare dispersion bandwdith {\it increases} with decreasing $\kappa$ [\figref{fig:sum}(b)] at $\theta=1.05^\circ$. In particular, for $\kappa=0.3$, the total width of the single-particle dispersion [panels (i),(j); red crosses] is $W\approx 14$ meV, almost twice as large as that of $\kappa=0.75$, explaining why in the former case the ground state is a semi-metal, whereas in the latter we obtain a gapped K-IVC state. 

The semi-metal state at $\kappa=0.3$ is still strongly correlated, however; only about half of the single-particle spectral weight is contained in the coherent quasi-particle dispersion [the sharp dispersing peaks in panel (j)]. The rest of the spectral weight is distributed over a wide energy range. At $K_M$ and $K'_M$ the spectral function exhibits broad peaks near the excitation energies of the interaction-only model [panel (j), blue crosses]. This spectral transfer is clearly seen in the total single-particle density of states, $\rho(\omega)$, shown in panel (h). At $T=58 \TK$, $\rho(\omega)$ exhibits two maxima centered around $\omega=\pm12\meV$. Below $T\sim 30 \TK$, these evolve into the broad `shoulders' around $\omega=\pm15 \meV$. The shoulders are distinct from the sharp maxima associated with the semi-metallic dispersion at $|\omega|\leq10 \meV$. 

We note the calculated $\rho(\omega)$ can be accessed in STM experiments, on averaging spectra  over the unit cell. The broad similarity between the experimentally observed spectra near charge neutrality in low temperature STM measurements eg. Ref.~\cite{Yazdani2019} and Figure \ref{fig:sum}(d) should be noted. More significantly, we further predict that the suppression of low energy spectral weight persists upto temperatures of $T \leq 10K$, beyond which it gradually fills in, which can be probed in future experiments.

\section{Twisted Bilayer Graphene}
\label{sec:TBG}

Before describing the QMC simulations, we first introduce the single-particle and many-body Hamiltonians for twisted bilayer graphene. 

\subsection{Single Particle Hamiltonian}
\label{sec:singleparticle}

In an appropriate basis, the continuum model single-particle Hamiltonian of TBG \cite{Bistritzer2011} can be written as follows (See \appref{app:B}):
\begin{align} \label{eq:tbg_ham}
    \hat{H}_\textrm{BM} &= \sum_{\bk}  c^\dagger_\bk H_\bk c_\bk + \sum_{\bk, \bk'} c^\dagger_{\bk'} T_{\bk',\bk}  c^{\,}_\bk   \nonumber \\
    H_\bk  &= \hbar v_F \qty( k_x \sigma_x \tau_z + k_y \sigma_y )
    \nonumber  \\
    T_{\bk',\bk} &= \frac{1}{2} \sum_n \delta_{\bk', \bk - \tau_z \bq_n}   \qty( \mu_x - i \mu_y ) T_n  + \textrm{h.c.} \nonumber\\
    T_n  &=  w_0 e^{-i \theta \sigma_z \mu_z \tau_z/2}  + w_1  e^{2\pi i n \sigma_z \tau_z /3} \sigma_x e^{-2\pi i n \sigma_z \tau_z /3}.
\end{align}
where $\theta$ is a twist angle, $v_F$ is the Dirac velocity, and $w_{0,1}$ is interlayer moir\'e hopping term. $c^\dagger_\bk$ represents a row vector of electron creation operators with wavevector $\bk$ and with sublattice, valley, layer, and spin indices, denoted by $(\sigma,\tau,\mu, s)$, respectively. Here $\bq_n = {\cal R}_{2\pi n/3}\, \bq_0$, where ${\cal R}_{\varphi} = e^{i \varphi \sigma_y}$ is a rotation matrix, $\bq_0 = \frac{8\pi \sin( \theta/2)}{3 \sqrt{3} a_0} (0,-1)^T$, and $a_0 = 1.42 \textrm{ \AA}$ is the intra-layer distance between carbon atoms. 

In the basis used here, the twist-angle dependence only appears in the $w_0$ term and in the moir\'e wavevector $\bm{q}_n$. Therefore, in the chiral limit \cite{Tarnopolsky} $w_0 \rightarrow 0$, the explicit twist-angle dependency in \eqnref{eq:tbg_ham} disappears. In this case, the Hamiltonian is invariant under the following many-body anti-unitary particle-hole transformation:
\begin{align}
    {\cal C} : \quad c^\dagger_\bk &\mapsto \tau_x \sigma_x \mu_y  c^{\,}_\bk \nonumber \\
     i &\mapsto -i 
     \label{eq:PHS}
\end{align}
such that, for $w_0=0$, $\cC \hat{H}_\textrm{BM} \cC^{-1} = \hat{H}_\textrm{BM}$. Note that if we denote the action of $\cC$ on the flavor indices as $U_\cC = \tau_x \sigma_x \mu_y$, the first-quantized Hamiltonian satisfies $U_\cC H_\bk U_\cC^{-1} = - H_\bk$ and $U_\cC T_{\bk,\bk'} U_\cC^{-1} = - T_{\bk,\bk'}$.

As we will demonstrate in Sec. \ref{sec:Sign}, the presence of an anti-unitary particle-hole symmetry is important for the absence of the sign problem in DQMC. However, away from the chiral limit, $w_0 \neq 0$, $\cC$ is weakly broken. The following \emph{small angle approximation} promotes the approximate PHS back to an exact symmetry by neglecting  the  rotation of the Pauli matrices in the individual Dirac Hamiltonians of the two graphene sheets,
\begin{equation}
    w_0 e^{-i \theta \sigma_z \mu_z \tau_z/2} \mapsto w_0\,.
\end{equation}
Since $\theta/\pi \ll 1$ near the magic angle, the wavefunctions and the combined bandwidth for conduction and valence bands barely change for $\kappa < 0.8$,  while the dispersion is now perfectly particle-hole symmetric unlike the original BM model Hamiltonian. (See \appref{app:B}).
As $\kappa \rightarrow 1$, the gap between the conduction band and the remote band starts closing and hybridizes, decreasing the overlap between flat band subspaces of the original BM model and approximated Hamiltonians. At this point, the idea of projecting the system into flat bands would not be valid anymore, and one has to include further neighboring bands to properly capture the hybridzation effect. However, in the experimentally relevant regime $\kappa$ is considered to be smaller than $0.8$ according to the first-principle calculations \cite{fang2019angledependent}. Therefore, the small angle approximation is justified, and will be used henceforth.

The resulting band structure with the correction from the interaction is shown in \figref{fig:sum}(a) and characterized by a pair of nearly flat bands at the Fermi level, while all other valence and conduction bands are separated by a large gap of order $15\meV$. Thus, the remote bands can be considered inert at low temperatures where valence (conduction) bands are completely filled (empty).
Foreshadowing the Coulomb interaction introduced in the next section, we project the charge-density operator, with respect to the background at charge neutrality, onto the nearly flat bands,
\begin{equation}\label{eq:denOps}
    \bar{\rho}_\bq = \sum_{\bk}  d^\dagger_{\bk + \bq} \Lambda(\bk,\bq) d^{\,}_{\bk} - \frac{1}{2} \sum_{\bk} \delta_{\bq, \bG}   \tr \Lambda(\bk,\bq)\,,
\end{equation}
where $d^\dagger_{\bk,i}$ is a creation operator for a Bloch eigenstate, $\ket{u_{\bk,i}}$, of flavour $i=(s,\tau,n)$, i.e., spin $s$, valley $\tau$, and band $n$. $d^\dagger_{\bk}$ is a row vector containing the eight operators $d^\dagger_{\bk,i}$. The momentum $\bk$ is restricted to the $1^{\mathrm{st}}$ moir\'e BZ while the transfer momentum $\bq$ remains unrestricted.
The form factor is defined as $\Lambda_{ij} (\bk,\bq) \equiv  \braket{u_{\bk+\bq,i}}{u_{\bk,j} }$.

The action the symmetry operators on $d_{\bk}$ depends on the choice of gauge. Particularly important for our purpose is the action of $\cC$, responsible for the absence of the sign problem. In \appref{app:C}, we describe a gauge fixing procedure such that
\begin{equation} \label{eq:PHS_projected}
    {\cal C} : \quad d^\dagger_\bk \mapsto \tau_x n_x  d^{\,}_\bk, \qquad     i \mapsto -i 
\end{equation}
where $n_{x,y,z}$ are Pauli matrices acting on the band index.

\subsection{Many-Body Hamiltonian}
\label{sec:many-body}

The parameters typically used for the BM model stem from first-principle calculations that include interaction effects, for example, the Dirac velocity is already renormalized. This implies that the QMC simulation has to avoid double counting \cite{Liu2021, BultnickKhalaf2020}. 
One approach to this subtle problem is to subtract the Hartree-Fock mean-field contribution of a properly chosen reference state, whose corresponding single-particle density matrix is given by $P_0$, as the following:
\begin{align}\label{eq:ham_QMC}
    \hat{H} &= \hat{H}_0  + \hat{V} \nonumber \\
    \hat{H}_0 &= \hat{H}_\textrm{BM} - \big[ \hat{V}\, \big]_{P_0}  \nonumber \\
    \hat{V} &= \frac{1}{2} \sum_\bq \bar{\rho}_\bq V_\bq  \bar{\rho}_{-\bq},
\end{align}
where $V_\bq = V_0 \tanh(\abs{\bq} d)/\abs{\bq}$ with $V_0 = \frac{e^2}{2 \epsilon \epsilon_0}\frac{1}{A}$ is the Fourier-transformed Coulomb potential with gate distance $d$ and relative permittivity $\epsilon$. Here, we use the density matrix with respect to the reference decoupled single graphene sheets, as discussed in \appref{app:D}. This is an intuitive choice since the first principle calculation were preformed for a monolayer graphene, however, this choice is not unique. In fact, one may also use the non-interacting ground state of the \emph{coupled} system as the reference state, and the resulting bandwidth can vary by a factor of $2$, apparent by comparing \cite{Liu2021} and \cite{BultnickKhalaf2020}.
Therefore, one should view the resulting corrected bare dispersion as an educated guess. Note that a particle-hole symmetric reference state $P_0$ induces a bare Hamiltonian, $\hat{H}_0$, that preserves the PHS $\cC$ and thus will not introduce a sign-problem.

\section{DQMC method for twisted bilayer graphene} \label{sec:Sign}

We now set up our problem for DQMC simulations, and show that the sign problem is absent at charge neutrality due to the PHS~\eqref{eq:PHS_projected}.

\subsection{Momentum-space algorithm and absence of the sign problem}
\label{sec:DQMC}

The algorithm we use to simulate the Hamiltonian~\eqref{eq:ham_QMC} is a variant of the Blankenbecler--Scalapino--Sugar algorithm~\cite{Blankenbecler1981}, adapted to work in momentum space~\cite{LL_SO5}. We first use a Trotter decomposition to write the partition function as
\begin{equation}
\label{eq:Z}
Z={\rm Tr}\left(e^{-\hat{H}_{0}\delta}\prod_{\bm{q},\alpha=1,2}e^{-\frac{V_{\bm{q}}}{8}\hat{\rho}_{\bm{q},\alpha}^{2}\delta}\right)^{N_{\tau}}+O(\delta^{2}).
\end{equation}
Here, the trace is taken over the many-body Hilbert space of fermions in the active bands, $N_\tau$ is the number of imaginary time steps, $\delta=\beta/N_\tau$ is the step size, and we have introduce the real and imaginary parts of the density operator:
\begin{align}
\hat{\rho}_{\bm{q},1}&=\bar{\rho}_{\bm{q}}+\bar{\rho}_{-\bm{q}},\nonumber\\
\hat{\rho}_{\bm{q},2}&=\frac{1}{i}\left(\bar{\rho}_{\bm{q}}-\bar{\rho}_{-\bm{q}}\right).
\end{align}
A subtlety that arises in Eq.~\eqref{eq:Z} is the ordering of the terms in the product over $\bm{q}$ and $\alpha$, since the $\hat{\rho}_{\bm{q},\alpha}$ operators do not commute with each other (due to the projection to the active bands). A symmetric ordering can be chosen in order to achieve a Trotter error that scales as $\delta^2$, as described in Appendix~\ref{app:QMCdetails}. For notational simplicity, we leave the ordering of the product implicit here.

We simulate a finite lattice of $L \times L$ moir\'e unit cells. Both crystal momentum $\bk$ and momentum transfer $\bq$ in Eq.~\eqref{eq:Z} are discretized as $(m_1 \bm{G}_1 + m_2 \bm{G}_2)/L$, where $G_{1,2}$ are moir\'e reciprocal lattice vectors and $m_{1,2} \in \mathbb{Z}$. Unlike $\bk$, $\bq$ is not restricted to the first Brillouin zone, and in principle has infinite range. However, for any sensible physical systems the magnitude of $V_\bq \bar{\rho}_\bq \bar{\rho}_{-\bq}$ decreases rapidly with $\bq$ as $\Lambda(\bk,\bq)$ decays rapidly with $\abs{\bq}$ \footnote{The main reason for the decay is the decay of the form factors contained inside $\bar{\rho}_\bq$ from the flat band projection.}, and one can set a cutoff for the range of $\bq$.  In our simulations, we restricted our $\bq$ to be within twice the size of the first Brillouin zone in each direction. 

To facilitate DQMC simulations, we perform a Hubbard-Stratonovich (HS) transformation to decouple the quartic terms in~\eqref{eq:Z}, writing the partition function as
\begin{equation}
Z = \sum_{\{\eta_{\bm{q},\alpha}(l)\}} W(\{\eta_{\bm{q},\alpha}(l)\})+O(\delta^{2}),
\end{equation}
where $\eta_{\bm{q},\alpha}(l)$ are real, discrete HS fields, and the weight $W(\{\eta_{\bm{q},\alpha}(l)\})$ is given by
\begin{equation}
\label{eq:W}
    W = {\rm Tr}\left[\prod_{l=1}^{N_{\tau}}\left(e^{-\hat{H}_{0}\delta}\prod_{\bm{q},\alpha}e^{-\hat{H}_{\rm{i}}(\bm{q},\alpha,l)\delta}f\left[\eta_{\bm{q},\alpha}(l)\right]\right)\right]. 
\end{equation}
Here, $\hat{H}_{\rm{i}}(\bm{q},\alpha,l)= i\left(\frac{V_{\bm{q}}}{8}\delta\right)^{1/2}\eta_{\bm{q},\alpha}(l)\hat{\rho}_{\bm{q},\alpha}$, 
$f[\eta_{\bm{q},\alpha}(l)]$ is a real, non-negative weight of the HS field, given explicitly in Eq.~\eqref{eq:eta_gamma_fields}. 

Proving that the sign problem is absent amounts to showing that $W(\{\eta_{\bm{q},\alpha}(l)\})$ is a real and non-negative for any configuration of the HS fields. To show this, we note that due to the $U(1)$ spin and valley conservation symmetries, we can write $\hat{H}_0 = \sum_{s,\tau}\hat{H}_{0,s,\tau}$, where $\hat{H}_{0,s,\tau}$ is the part of $\hat{H}_0$ that acts on electrons with spin $s$ and valley $\tau$, and similarly $\hat{H}_{\rm{i}} = \sum_{s,\tau}\hat{H}_{\rm{i},s,\tau}$. Using this fact, we can decompose $W$ as a product: \begin{equation}
    W(\{\eta_{\bm{q},\alpha}(l)\})=\prod_{s,\tau}W_{s,\tau}(\{\eta_{\bm{q},\alpha}(l)\}), 
\end{equation}
where $W_{s,\tau}$ is of the form of Eq.~\eqref{eq:W}, replacing $\hat{H}_{0}$ by $\hat{H}_{0,s,\tau}$ (and similarly for $\hat{H}_{\rm{i}}$), and taking the trace over electrons with spin and valley indices $(s,\tau)$.

We now use the symmetry of the Hamiltonian under $\mathcal{C}$ [Eq.~\eqref{eq:PHS_projected}]. In particular, $\mathcal{C}\hat{H}_{0,s,\tau} \mathcal{C}^{-1} = \hat{H}_{0,s,-\tau}$, and similarly for $\hat{H}_{\rm{i},s,\tau}$. Using this transformation law, and accounting for the anti-unitarity of $\mathcal{C}$, one can show that (see Appendix~\ref{app:A}, \eqnref{eq:anti-unitary_theorem_app})
\begin{equation}
    W_{s,\tau}(\{\eta_{\bm{q},\alpha}(l)\})=W_{s,-\tau}^{*}(\{\eta_{\bm{q},\alpha}(l)\}).
    \label{eq:Wst}
\end{equation}
Hence, we find that $W(\{\eta_{\bm{q},\alpha}(l)\})=\prod_{s}\left|W_{s,\tau}\right|^{2}\ge0$, and there is no sign problem. As a corollary, we note that the weight of a single spin flavor is real and non-negative on its own. Therefore, a spin-polarized system with a single spin flavor is also sign problem free.

The DQMC algorithm proceeds by interpreting $W/Z$ as the probability of the configuration $\{\eta_{\bm{q},\alpha}(l) \}$, and sample the configuration space using a Markov Chain Monte Carlo method. Since both $\hat{H}_0$ and $\hat{H}_{\rm{i}}$ are quadratic in the fermion operators, the fermionic trace can be performed exactly for a given configuration of the HS fields (\appref{app:QMCdetails}).

\begin{table}[t]
\setlength\tabcolsep{5 pt}
\renewcommand{\arraystretch}{1.3}%
\begin{tabular}{|c|c|c|c|}
 \cline{2-4}
\multicolumn{1}{c|}{}  & Chiral & Nonchiral & Intervalley  \\
\hline
$\mqty{ \textrm{Required}  \\ \textrm{Multiplicity} }$ & $\mqty{ \textrm{2 valleys}  \\ \textrm{OR 2 spins} }$& 2 valleys & $\mqty{ \textrm{4 valleys}  \\ \textrm{AND spins} }$ \\
\hline
\end{tabular}
\vspace{0.05in}
\caption{\label{tab:signfree} The minimal number of bands required to guarantee the absence of sign problem in DQMC.  We always assume a conduction/valence band pair, and note the multiplicities of such pairs to obtain a sign problem free formulation. For the nonchiral case ($\kappa > 0$), we always need at least {\em two} conduction-valence band pairs in opposite valleys, since it is the valley-exchanging anti-unitary symmetry that guarantees the absence of sign problem. However, with chiral symmetry ($\kappa=0$), even in the presence of band dispersion, a single valley is sign problem free, provided both spin-species are present. Finally, we note that if we include intervalley density-density interactions, that are typically suppressed at small angle, the HS-decoupled Hamiltonian is not block-diagonal in valley anymore and we need all four pairs of bands (both spin and valley) to guarantee the absence of sign problem.}
\end{table}

\subsection{Sign-problem-free extensions}

Employing the identity \eqnref{eq:anti-unitary_theorem_app}, we now consider various extensions and physical perturbations to the TBG Hamiltonian~\eqref{eq:ham_QMC} that allow for sign problem-free DQMC simulations. Table~\ref{tab:signfree} summarizes different cases of the TBG where the sign problem is absent, labeled by whether we are working in the chiral limit ($\kappa=0$) or not, and whether an inter-valley scattering interaction is included [see Eq.~\eqref{eq:rho_inter}]. For each case, we specify the number of conserved flavors (valley and spin) needed to avoid a sign problem.

As we have seen, in the most generic case (away from the chiral limit, $\kappa>0$, and with $\hat{H}_0\ne 0$), the DQMC weight associated with each spin flavor is real and non-negative, and therefore a problem with a single spin flavor is sign problem-free [Eq.~\eqref{eq:Wst} and the discussion below].  What if we have more symmetries? In the chiral limit ($\kappa = 0$), we can relax the condition for the sign problem further. In this case, we have an extra anti-unitary particle-hole symmetry $\cC_\textrm{sub}$ whose unitary part is simply given as $U_{\cC_\textrm{sub}} = \tau_z n_x$ in the band basis (See \tabref{tab:sym_rep}). The origin of $\cC_\textrm{sub}$ is the sublattice symmetry of the original graphene layer. Since it maps each part of the HS-decoupled Hamiltonian with a given $(s,\tau)$ onto itself, $W_{s,\tau}$ must be real. For a single valley, if we have both spin flavors, then the presence of the unitary spin-flip symmetry $s_x$ implies that $W_{s,\tau} = W_{-s,\tau}$ due to \eqnref{eq:anti-unitary_theorem_app}. Therefore, $W_\tau \equiv W_{\uparrow,\tau} W_{\downarrow,\tau} \geq 0$. In other words, one can simulate a single valley of TBG without a sign problem in this case. 

In the special case where the bare dispersion $\hat{H}_0$ is set to zero in the chiral limit, we can show that only two bands related by $\cC$ symmetry are needed to guarantee the absence of sign problem. Here we can choose a basis which diagonalizes $n_x$, denoted by $\tilde{n}$ (\appref{app:C}). Then, HS-decoupled Hamiltonian is diagonal in $(s,\tau,\tilde{n})$. Furthermore, the extra anti-unitary particle-hole symmetry $\cC_\textrm{sub}$ maps each flavor to itself since it commuts with $n_x$. It implies that the corresponding determinant $M_{s,\tau,\tilde{n}}$ is real. Now, note that the unitary part of $\cC$ in this basis is given as $U_{\cC} = \tau_y \sigma_y$. It implies that $W_{s,\tau,\tilde{n}} = W_{s,-\tau,-\tilde{n}}^*$. Therefore, $W_{s,C} \equiv W_{s,\tau,\tilde{n}} W_{s,-\tau,-\tilde{n}}$ is real and non-negative, where $C = \tau_z \sigma_z$ is the Chern number for this pair of flavors.

Finally, it is worth noting that one can add an inter-valley density interaction. Such an interaction was not considered in our current simulations, due to the fact that it is estimated to be smaller than the long-range Coulomb interaction by a factor of $a_0/a_M \sim \sin(\theta)$, i.e. the ratio of the atomic lattice constant $a_0$ to the  moir\'e lattice constant. $a_M$.  However, this interaction is allowed by symmetry, and despite its small magnitude it can have significant effects in lifting degeneracy. The inter-valley interaction is of the form $\hat{H}_{\rm{iv}} \propto \sum_{\bm{q}}\rho_\bq^{\textrm{inter}}\rho_{-\bq}^{\textrm{inter}}$, where the inter-valley density is defined as
\begin{equation}
\label{eq:rho_inter}
    \rho_\bq^{\textrm{inter}} = \sum_\bk d_{\bk+\bq}^\dagger \Lambda^{\textrm{inter}}(\bk,\bq) d_{\bk}^{\,}. 
\end{equation}
$\Lambda^{\textrm{inter}}(\bk,\bq)$ is off-diagonal in  valley space. This interaction reduces the manifold for the K-IVC state from U(2) to U(1) for spin singlet K-IVC case, which allows for a finite-temperature transition. Once this interaction is introduced, the HS-decoupled Hamiltonian is not block-diagonal in valley anymore. As a result, using $\cC$ symmetry, we can show that $W_{s}$ is real, but not necessarily non-negative. By further using spin-flip unitary symmetry $s_x$, we can show that $W_\uparrow W_\downarrow > 0$. Therefore, we need all physical flavors of TBG to be present in this case to evade the sign problem.

Using our approach, we can further show that the model remains sign-problem free in the presence of external fields~\footnote{First quantized matrix for such operator should have zero trace, unless it is chemical potential}. First, consider the Zeeman field term
\begin{equation}
    \hat{H}_\textrm{Z} = -h_z \sum_{\bm{k}} d^\dagger_\bk s_z d^{\,}_\bk.
\end{equation}
Let $\hat{H}_s$ be a HS-decoupled Hamiltonian for spin flavor $s$, containing both valleys and bands. Then, we can show that $(s_x \cC) \hat{H}_{s} (s_x \cC) = \hat{H}_{-s}$. Therefore, using anti-unitary particle-hole symmetry $s_x \cC$ we obtain that  $W_{\uparrow}^* = W_{\downarrow}$ and $W = W_{\uparrow} W_{\downarrow} = |W_{\uparrow} |^2 > 0$. 

We can also include a `valley-Zeeman' field,
\begin{equation}
    \hat{H}_\textrm{vZ}=  -h_{vz} \sum_{\bm{k}}d^\dagger_\bk \tau_z  d^{\,}_\bk.
\end{equation}
This term is symmetric under $\cC$, and does not cause a sign problem even if we have a single spin flavor. Similarly, we can add the uniform sublattice polarization ($\sim \tau_z n_x$ in band basis) and perpendicular electric field terms, which are also symmetric under $\cC$. Therefore, in principle we can turn on all these four kinds of external fields and simulate the system without a sign problem. 

\section{Simulation Results and Discussion}
\label{sec:result}
In this section we report the results of our QMC computations. Our implementation of the algorithm described above is built on top of the software package ALF \cite{ALF2017,ALF2021}.
We investigate the leading instabilities of the BM model at charge neutrality, and benchmark our simulations by comparing the results to the analytical solutions in the strong coupling limit \cite{BultnickKhalaf2020}.
Furthermore, we present a more detailed analysis of the spectral functions shown in \figref{fig:sum} focusing on their temperature dependence.

\subsection{Parameters}
\label{sec:parameters}
We used to following set of parameters in the simulations.
In the BM model, we used $\frac{\hbar v_F}{3a_0/2} = 2700\,\textrm{meV}$ and $w_1 = 105\,\textrm{meV}$. We set $\theta = 1.05$ which is close to the magic angle for the above parameters. For the interaction, we focus on a relative dielectric constant $\epsilon=10$ and a gate distance of $d=20\,\mathrm{nm}$. We confirmed the convergence of QMC results for a Trotter step size of down to $\delta=0.01\,\textrm{meV}^{-1}$ as shown in \figref{fig:flat}.

Effectively, two most important parameters in the simulation is the bandwidth $W$ of bare dispersion $\hat{H}_0$ and the interaction energy scale. It is worth noting that $W$ is a complicated function of the system parameters. Due to the correction term in \eqnref{eq:ham_QMC}, it depends on the interaction parameters, $\epsilon$ and $d$, as well as on the choice of reference state $P_0$ and the hopping ratio $\kappa = w_0/w_1$. The interaction energy scale can be represented by the bandwidth of electron and hole excitations in the strong coupling limit, $\hat{H}_0=0$ (\appref{app:D}). In \figref{fig:sum}(b), we plot $W$ and the interaction scale as a function of $\kappa$ at $\theta = 1.05^\circ$. In this plot, $W$ decreases with $\kappa$, while the interaction scale barely changes. However, this behavior is not universal; as we move away from the magic angle, $W$ can increase with $\kappa$. (\figref{fig:different_angle}).

\subsection{Observables}

In our DQMC simulation, there are two observables we examine: ($i$) the electron Green's function $G(\bk, \tau)$, and ($ii$) correlation functions $S_a$ of fermion bilinears ${\cal O}_a$,
\begin{align}
    \label{eq:order_parameter} 
    S_a(\bq) &\equiv \frac{1}{L^2}\expval{ {\cal O}_a(\bq) {\cal O}_a(-\bq) } \nonumber \\
    {\cal O}_a(\bq) &\equiv  \sum_\bk d^\dagger_{\bk + \bq} \Lambda_a(\bk, \bq) d^{\,}_\bk.
\end{align}
$N_M = L^2$ is the number of moir\'e unit cells in the system and $\Lambda_a$ is the operator-dependent form factor,
\begin{equation}
    \qty[ \Lambda_a(\bk,\bq) ]_{ij} \equiv  \bra{u_{\bk+\bq, i}} M_a \ket{u_{\bk, j}},
\end{equation}
where $i,j$ runs over the flavor indices $(s,\tau,n)$ of the projected flat bands. The microscopic operator $M_a$ for various correlation functions are defined in \tabref{tab:OP}, which act on the vector space of the unprojected electron operator. 
The normalization of \eqref{eq:order_parameter}  implies that a spatially homogeneous long-range order gives rise to an extensive scaling of $S_a(\bq=0)\sim L^2$.
Note that the HS decoupled Hamiltonian is a fermion bilinear such that higher correlation function can be extracted using Wick's theorem, as illustrated in \appref{app:C}\,4. 
The correlation length, $\xi_{a}$, which we presented in \figref{fig:sum}, is extracted from the momentum dependence of $S_a(\bq)$ at small wavevectors \cite{Toldin15},
\begin{equation}
\xi^2_{a}/a^2_M=\frac{3}{16\sin^2(\pi / L)} \left(\frac{S_a(\bq=0)}{S_a(\bq=\bq_{nn})}-1\right),\label{eq:CorrLength}
\end{equation}
where $\bq_{nn}$ are the six nearest-neighboring momenta of $\bq=0$, $S_a(\bq=\bq_{nn})$ is averaged over them.

At each momentum $\bk$ we have eight electron flavors, and therefore there is a large manifold for potential symmetry breaking. In order to examine the nature of the ground state manifold, we choose six representative order parameters: valley polarized (VP), spin polarized (SP), valley-Hall (VH), two types of intervalley coherence (t-IVC/K-IVC), and quantum Hall (QH), whose correspodning microscopic operators $M_a$ are listed in \tabref{tab:OP}.

\begin{figure*}[t]
    \centering
    \includegraphics[width = 1 \textwidth]{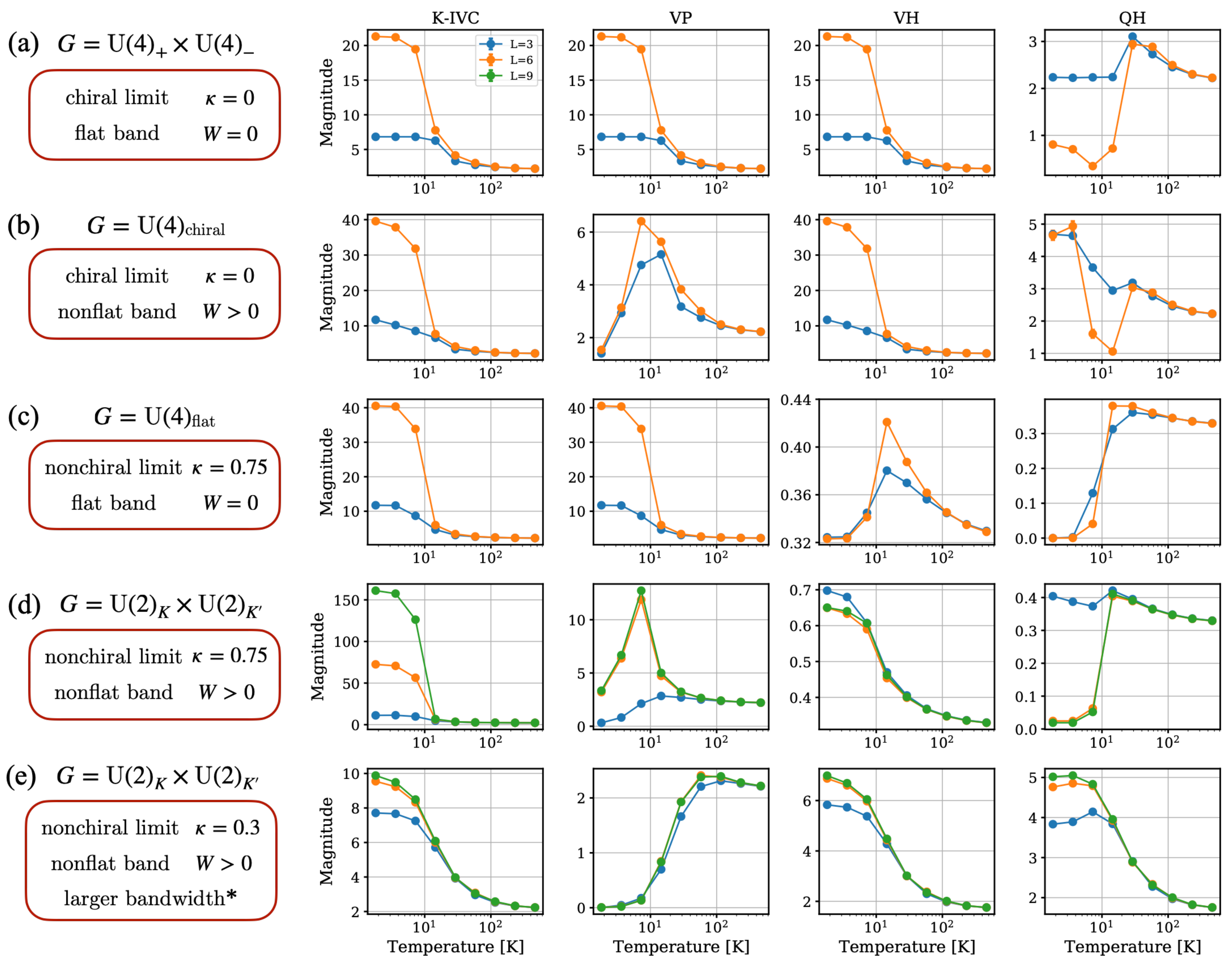}
    \vspace{-0.1in}
    \caption{\label{fig:QMC_channel} Correlation functions at $\bq = 0$ and global symmetry $G$ of full-interacting TBG Hamiltonians in four different limits ($\epsilon=10$ and $d_s=20$nm). Each limit is characterized by: $(\kappa,W)$ where $\kappa = w_0/w_1$ and $W$ is the bandwidth of the bare dispersion $\hat{H}_0$ in \eqnref{eq:ham_QMC}. (a) chiral, flat limit: $(0,0\meV)$, (b) chiral, non-flat with artifically reduced bandwidth: $(0,3.8\meV)$, (c) non-chiral, flat: $(0.75,0\meV)$, (d) non-chiral, non-flat: $(0.75,8.3\meV)$, and (e) non-chiral, non-flat: $(0.3,14.3\meV)$.
     Correlation functions  $S_i(\bq = 0)$ for K-IVC, VP, VH and QH are shown, which should scale as the system volume if they acquire long-range order.
    In (a), all channels have equal magnitude of correlation functions. The suppression of $S_{\textrm{QH}}(\bq=0)$ in this case is due to the difference in the degeneracy of $C_\textrm{tot}=0$ and $C_\textrm{tot}=2$ groundstate manifold sectors. The deviations from this ideal limit of $(\kappa,W)=(0,0\meV)$ lift the degeneracy among symmetry broken states, as tabulated in \tabref{tab:OP}. In (b) where we retain a  quarter of the bare dispersion to obtain an ordered state, the $U(4)_{\rm chiral}$ relates the K-IVC and VH orders, while the VP is suppressed due to dispersion. 
    In (c), compared to the chiral nonflat case, the system now has a different $\U(4)_\textrm{flat}$ symmetry which relates K-IVC and VP states instead of VH. Here, VH is suppressed due to nonchiral part of the interaction.
   Progressing from (a) to (d), each perturbation reduces the symmetry down to that in the realistic limit, $\U(2)_\bK \times \U(2)_{\bK'}$, where only the K-IVC groundstate survives. \PRX{In (e), we are in the same symmetry class as in (d), but a larger bandwidth. As a result, the correlation functions saturate to a finite value with increasing system size, resulting in a disordered phase. In (e), although not shown, we emphasize that the magnitudes of both VP and SP correlations are vanishing at low temperature. }} 
\end{figure*}

\begin{table}[t]
\centering
\setlength\tabcolsep{4pt}
\renewcommand{\arraystretch}{1.15}%
\begin{tabular}{|c|c|c|c|c|c|}
 \cline{2-6}
    \multicolumn{1}{c|}{}  & $\mqty{ \kappa = 0 \\\, W = 0 \,}$ & $\mqty{ \kappa = 0 \\ \, W > 0 \,}$ & $\mqty{ \kappa > 0 \\ \,W = 0 \,}$ & $\mqty{ \kappa > 0 \\\, W > 0 \,}$  & $M_i$  \\
\hline 
VP &  \cmark & \xmark & \cmark & \xmark & $\tau_z$\\
\hline
SP &  \cmark & \xmark & \cmark & \xmark & $s_{z}$ \\
\hline
QH & \cmark & \cmark & \xmark & \xmark & $\tau_z {\sigma}_z$ \\
\hline
VH & \cmark & \cmark & \xmark & \xmark & $ {\sigma}_z$ \\
\hline
t-IVC& \cmark & \xmark & \xmark & \xmark & $\tau_{x,y} \sigma_y \mu_y$  \\
\hline
K-IVC& \cmark & \cmark & \cmark & \cmark & $\tau_{x,y} \sigma_x \mu_y$   \\
\hline

\end{tabular}
\vspace{0.05in}
\caption{\label{tab:OP} Order parameters and regimes where the  corresponding symmetry broken phases are favored for the parameters $(\kappa, W)$ in the strong coupling limit \cite{BultnickKhalaf2020}, i.e., small dispersion limit. In the last column, microscopic operators are listed for each order parameter.
In the chiral limit with dispersionless (flat) band, i.e., $(\kappa,W) = (0,0)$, the system has $\U(4) \times \U(4)$ global symmetry \cite{BultnickKhalaf2020}. Accordingly, symmetry broken states for the listed order parameters become degenerate.
For $W > 0$, the finite dispersion gives rise to an anti-ferromagnetic coupling between different Chern sectors and disfavors the SP, VP, and t-IVC states. 
For $\kappa > 0$, the deviation from chiral limit gives rise to 
inter-sublattice interactions which disfavors QH, VH, and t-IVC states. 
Combining both perturbations, the K-IVC becomes the most favored state among the listed symmetry broken states.
This qualitative behavior agrees with the observations in \figref{fig:QMC_channel}    }
\end{table}

\subsection{Hierarchy of Symmetries and Degeneracy of Ground-state Manifold}\label{sec:GShierarchy}

In the following, we expose the hierarchy of scales and symmetries in TBG by studying the temperature dependence of various correlation functions in different limits.
For chiral and nonchiral limits, we have $\kappa = 0$ and $0.75$ respectively. 
For flat and nonflat bands, we used $\hat{H}_0 = 0$ and $\hat{H}_0 = \alpha (\hat{H}_\textrm{BM} - \big[ \hat{V}\, \big]_{P_0})$ respectively in  \eqnref{eq:ham_QMC}. For the nonchiral limit, we take $\alpha=1$  corresponding to a bandwidth of $W=8.3 \meV$ at $\kappa = 0.75$ (realistic limit). For the chiral limit $\kappa = 0$, we suppress the dispersion with $\alpha=0.25$ so the bandwidth $W^*=3.8\meV$ stays within the ordered phase and allows a meaningful  comparison. $\alpha=1$ leads to a correlated semimetal, that we study using $\kappa=0.3$ with $W=14.3 \meV$. 
In each limit, we have different global symmetries as summarized in \figref{fig:QMC_channel}.
Therefore, correlation functions must exhibit an exact degeneracy induced by the global symmetry.  For a detailed symmetry analysis, we refer to \appref{app:sym_analysis}.
The results for the correlation function $S_a(\bq=0)$ are depicted in \figref{fig:QMC_channel} as function of temperature for four different classes: (a) flat (i.e., $\hat{H}_0=0$) chiral, (b)  nonflat chiral, (c) flat nonchiral, and  (d,e) nonflat nonchiral. Note that (d) and (e) have the same global symmetry, but (e) has a larger bandwith.
For readability, we omit the correlation functions for SP and t-IVC order parameters. The magnitude of SP is equivalent to that of VP by symmetry. The magnitude of t-IVC is equivalent to the others only at chiral flat limit, and is suppressed in all other cases.   
\PRX{\figref{fig:QMC_channel}(d),(e) demonstrate that all short-ranged correlation functions converged with system size while $S_{\mathrm{K-IVC}}(q=0)$ for $\kappa=0.75$ [\figref{fig:QMC_channel}(a)-(d)] systematically increases with system size at low temperature.  In fact, $S_{\mathrm{K-IVC}}(q=0)/L^2$ determines the K-IVC order parameter at zero temperature by first taking the limit $T\rightarrow 0$ and then extrapolating to the thermodynamic limit.}

The correlation function saturates as we decrease the temperature to a value that scales with the system's volume for the $L=3,\,6$, as shown in Fig. \ref{fig:QMC_channel}. Based on this observation, we can conclude that at these temperatures and system sizes, the system appears to be long-range ordered. However, it should be emphasized that no true long-range or even algebraic order is expected at finite temperature in thermodynamic limit, as the order parameter break a continuous non-Abelian symmetry.

The data confirms the theoretical understanding of the hierarchy of symmetries in TBG. In the flat chiral limit, conduction and valence bands are degenerate and the form factors for the Coulomb interaction obey an enhanced symmetry (see \appref{app:sym_analysis}). As a result, the system exhibits a $\U(4)_+ \times \U(4)_-$ symmetry, where each $\U(4)_\pm$ acts on the respective Chern sector $C = n_x$ in the band basis ($\tau_z \sigma_z$ in a sublattice basis of \cite{BultnickKhalaf2020}). 
Within each Chern sector, there are four electron flavors (valley index $\tau=\pm 1$ and spin index $s = \pm 1$), and the system is invariant under rotations of those flavor.
Due to this large symmetry group, there is a massive ground state degeneracy at charge neutrality. All the ground states are annihilated by the Coulomb interaction. 
The ground states reside in the union of following disjoint manifolds in the thermodynamic limit (See \appref{app:degeneracy}):
\begin{align} \label{eq:gs_manifold}
    \qty(\frac{\U(4)}{\U(2)\times \U(2)})^{\otimes 2}  \bigcup \qty(\frac{\U(4)}{\U(3)\times \U(1)})^{\otimes 2}   \bigcup\,   \mathbb{Z}_2 
\end{align}
where the three manifolds correspond to the states with Chern number 0, $\pm2$, and $\pm4$. 

Such a disjoint set of ground state manifolds indicates that the system is fine-tuned to a first order phase transition. This situation is typically challenging for the DQMC algorithm due to the breakdown of ergodicity. Note that the PHS symmetry which guarantees the absence of the sign problem is preserved by each configuration and therefore enforces an equal sampling of opposite Chern sectors, $\pm C$. This already facilitates the sampling. Additionally, we have carefully analyzed QMC runs and fine-tuned updates, e.g. by using both discrete and continuous auxiliary fields for the $\bq=0$ contribution of the Coulomb repulsion. The results presented in \figref{fig:sum} are taken at finite $\kappa$ where this potential issue is absent.

Let us note a few consequences arising from the different dimensions of the ground state manifolds. In a finite-size system with $N_M = L^2$ moir\'e unit cells, one can calculate the degeneracy of each manifold using Young's tableau. The calculation in \appref{app:degeneracy} shows that while the degeneracy of $C=0$ ground states scales as $N_M^8$, the degeneracy of $|C|=2$ ground states scales as $N_M^6$. For $|C|=4$, there are only two ground states.
Hence, the $C=0$ sector (containing e.g. K-IVC, t-IVC, SP, VP, VH states) dominates the partition function.
Indeed, if we look at the QH channel (corresponding to $|C|=2,\,4$) in \figref{fig:QMC_channel}(a), the QH correlation function decreases with the system size while the other channels (K-IVC, VP, VH, etc.) increase. This aligns with theoretical expectation: the relative scaling of the degeneracy between $C=0$ and $\abs{C}=2$ scales as $1/N_M^2$  (the $\abs{C}=4$ only contributes 2 states and can be neglected).  Therefore the QH correlation function at $\bq = 0$ should scale as $N_M \cdot 1/N_M^2 = 1/N_M$, where the first factor of $N_M$ is the volume scaling. This accounts for the suppression of the QH channel at low temperature. Nevertheless, the correlation function is strongly peaked at $\bq=0$ giving rise to a very large correlation length (not shown).

As we turn on the dispersion, the perturbation induces anti-ferromagnetic like interaction among local order parameters \cite{BultnickKhalaf2020} that disfavors the t-IVC, SP, and VP ground states, as illustrated in \tabref{tab:OP}. As a result, the groundstate manifold shrinks and the saturation value of the order parameter magnitudes should increase, which agrees with our observation in \figref{fig:QMC_channel}(b). 
One important caveat is that an anti-ferromagnetic interaction favors totally singlet ground-states. For example, on a finite-size lattice, the Heisenberg antiferromagnet has a unique groundstate, unlike the ferromagnet whose ground state degeneracy scales as the system's volume. However, once we consider the Anderson tower of states within a small energy window $[0,1/\beta]$ set by finite temperature, the number of states inside the window rouhgly scales with $N_M$ as if the order parameter belonged in the following manifold:
\begin{align} \label{eq:gs_manifold2}
    \qty(\frac{\U(4)}{\U(2)\times \U(2)})   \bigcup \qty(\frac{\U(4)}{\U(3)\times \U(1)})    \bigcup\,   \mathbb{Z}_2 
\end{align}
Accordingly, the quasi-degeneracy of each manifold with a different Chern number roughly scales as $N_M^4$, $N_M^3$, and $2$, respectively. For a thorough discussion, see \appref{app:degeneracy}. This implies that QH correlation functions should now scale as $N_M \cdot 1/N_M \sim \textrm{Const}$, which aligns with our observation in \figref{fig:QMC_channel}(b) where $S_\textrm{QH}(\bq=0)$ at different system sizes collapse at low temperature.
Note that this scaling behavior becomes asymptotically exact at thermodynamic limit for any finite temperature $T>0$ since the energy spacing between Anderson tower states scales as the inverse of system's volume. At zero temperature or temperature scale far below the excitation gap of Anderson tower states, $C=0$, $C=\pm 2$, and $C=\pm 4$ sectors will collapse into unique ground states respectively. Therefore, the QH correlation function should exhibit a volume-law scaling behavior. 
However, a rough estimate of the Anderson tower excitation energy for $L=6$ shows that this limit is reached when $T\sim 0.2\,\rm{K}$, which is lower than the temperature in our DQMC simulations.

We can use the results of \figref{fig:sum} to determine the stiffness of the non-linear sigma model describing the ground state manifold~\cite{KhalafSC, SoftModes}. This value controls the dispersion of the low-energy charge neutral bosonic modes \cite{SoftModes} as well as skyrmion excitations \cite{KhalafSC}. As shown in \appref{App:correlationLength}, the stiffness $\rho$ can be extracted using the correlation length data. 
By fitting the relation $\xi \sim e^{\pi \rho/2 T}$ which is valid above the saturation temperature \figref{fig:stiffness}, we extract $\rho \sim 1 \meV$.  The value of $\rho$ obtained here is in rough agreement with the value obtained through Hartree-Fock calculation in Ref.~\cite{KhalafSC}.

\begin{figure}[t]
    \centering
    \includegraphics[width = 0.49 \textwidth]{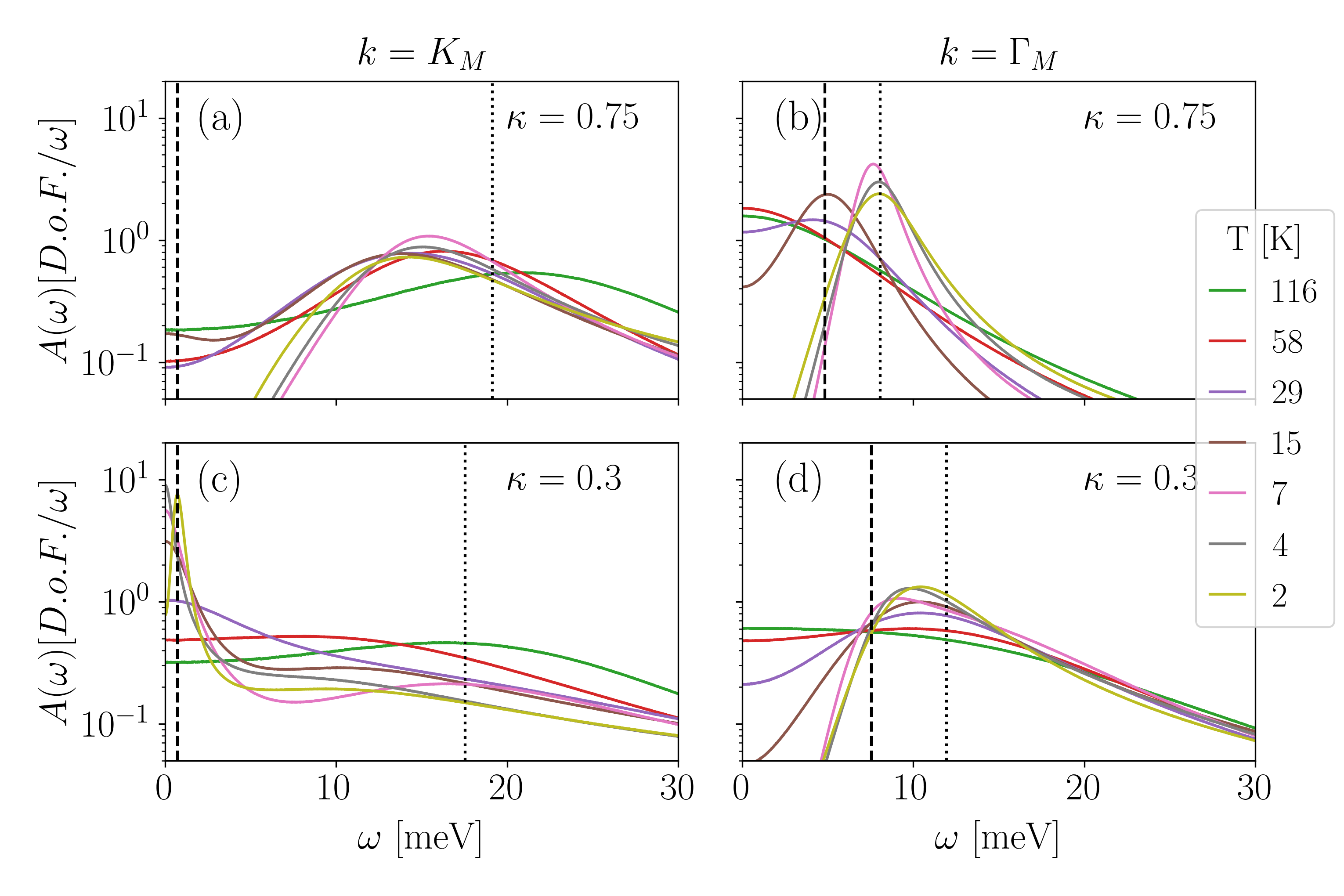}
    \caption{
    Spectral function $A(\omega,k)$, summed over band, valley and spin, at the Dirac point $\bk=\KM_M$ (a,c) and the BZ center $\bk=\Gamma_M$ (b,d) for $\kappa=0.75$ (top row) and $\kappa=0.3$ (bottow row) and $L=9\times 9$. The dashed vertical lines indicate the non-interacting excitation energy, shifted by the charging energy, the dotted line represent interaction-only limit. We find that the spectra for $\kappa=0.75$ are in fairly good agreement with the excitation energies of the interaction-only limit. For $\kappa=0.3$, instead, sharp peaks arise at low temperatures at the Dirac point $\bk=\KM_M$, their energy agrees with the finite-size charging energy, indicating the semi-metallic phase. Note the spectral weight at $\omega\sim \pm 15\meV$, reminiscent of the purely interacting mode. One cannot distinquish those two modes at $\bk=\Gamma$.  \label{fig:SpecCut}    }
    
\end{figure}

\subsection{Spectral Function and Charge Gap}

The momentum dependence of the spectral function has already been shown in \figref{fig:sum} and has been discussed in \secref{sec:sum}. Let us therefore focus on the frequency and temperature dependence of spectral function at particular momenta, presented in \figref{fig:SpecCut}. The spectral function is reconstructed using maximum entropy method \cite{MaxEnt2004} from the electron Green's function $G(\tau)$, shown in \figref{fig:Gtau} of \appref{app:QMCdetails}.

{\bf Spectral evolution with temperature in the K-IVC insulator:} In \figref{fig:SpecCut}(a), we show the spectral function $A(\omega,\bm{k}=\KM_M)$ (at the Dirac point) for $\kappa=0.75$ at different temperatures. At the lowest temperatures, we observe two clear peaks located around $\pm 19 \meV$, whose values are very close to the exact excitation energies of the interaction only Hamiltonian ($H_0=0$), which also has a K-IVC ground state~\cite{BultnickKhalaf2020}. Clearly, since the $H_0$ has gapless excitations at the Dirac points, the observed gap is a result of interactions. 

As we increase the temperature, the K-IVC correlation length become smaller than one lattice spacing around $10\TK$, as shown in \figref{fig:sum}(c) and \figref{fig:QMC_channel}(d). Surprisingly, this does not immediately lead to the disappearance of the gap-like features in the spectral function. Rather, the location of the peak in $A(\omega,\bm{k}=\KM_M)$ broadens and moves towards $\pm 15 \meV$. Such peaks persist for an intermediate range $10\,\textrm{K} \lesssim T \lesssim 100\,\textrm{K}$. The spectral weight at $\omega=0$ increases with increasing $T$, which may arise as the superposed tails of two thermally broadened peaks. By $T=116\TK$, the gap finally fills in, and there is a single, broad peak centered at $\omega=0$.

The spectrum at the $\Gamma_M$-point is shown in \figref{fig:SpecCut}(b). The picture is qualitatively similar to that at the $K_M$ point, although the magnitude of the gap is smaller. The gap features fills in by $T=60\TK$. 

To summarize the results for $\kappa=0.75$, the spectra show a pseudo-gapped regime at intermediate temperatures before a fully gapped state with substantially long-ranged K-IVC correlations develops below $10\,\rm{K}$. The energy density obtained by the DQMC simulation for $L=6$ at $T=1.8 \TK$ is $E=-0.377 \pm 0.006 \meV$ per moir\'e unit cell, which is in good agreement with the self-consistent Hartree-Fock calculation result, $E=-0.384 \meV$ per moir\'e unit cell. Furthermore, the excitation spectrum obtained in the DQMC coincide with the HF solution. This indicates that the Hartree-Fock calculation is accurate in this case, i.e., the ground state is well approximated by a Slater determinant.

\begin{figure}[t]
    \centering
    \includegraphics[width = 0.5 \textwidth]{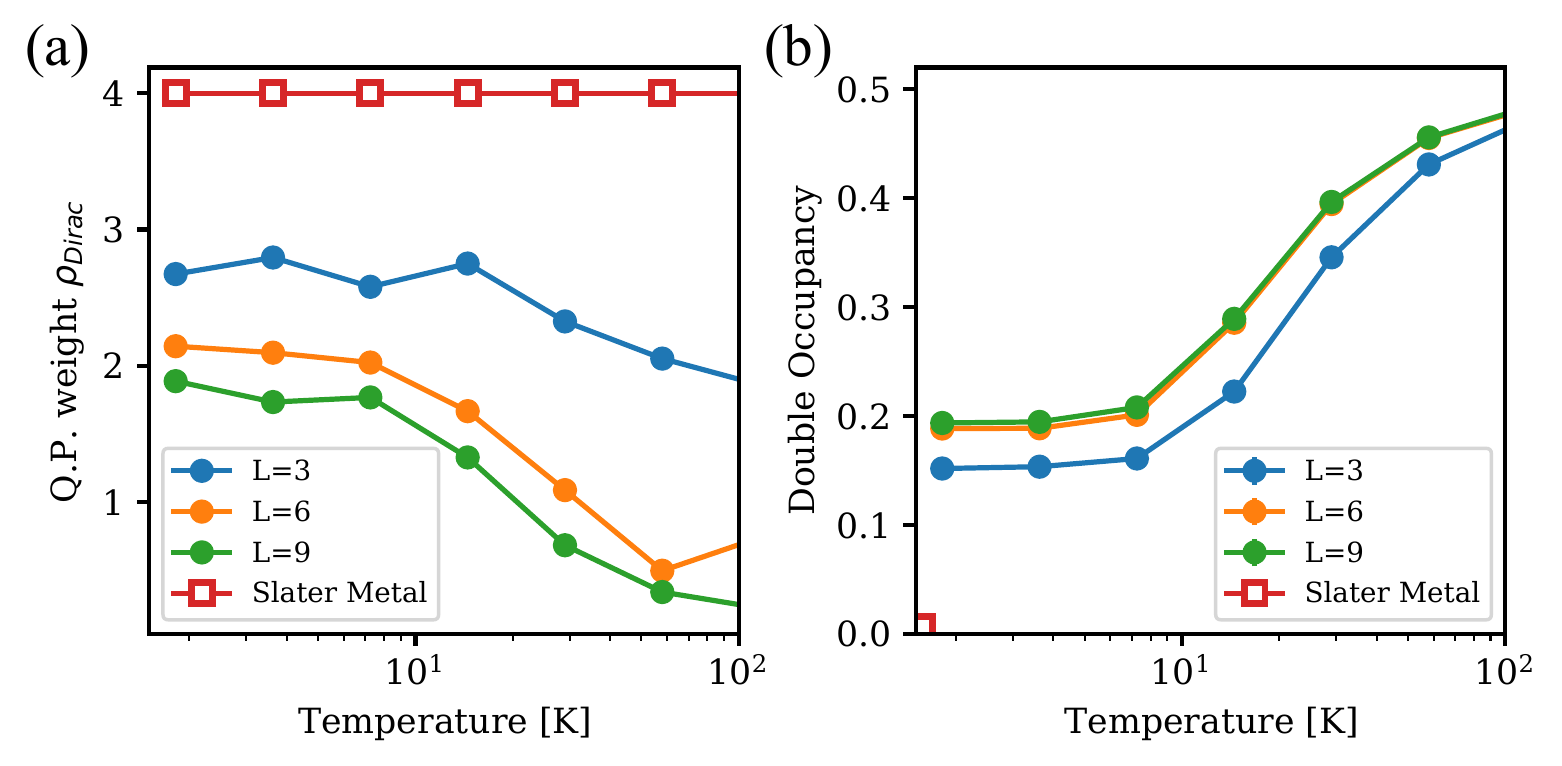}
    \caption{\PRX{Characterization of the strongly correlated semi-metal phase at  $\kappa=0.3$. (a)
    Proxy for the quasiparticle weight $Z$ of gapless Dirac particle in comparison to the non-interacting Dirac metal.
    (b) Flavor resolved double occupancy averaged over all spin, valley, and momenta indices. 
    }\label{fig:semimetal}
    }
    
\end{figure}

 {\bf Spectral evolution with temperature in the semi-metal:} The second row of \figref{fig:SpecCut}(c,d) shows the same quantities as (a,b), but for $\kappa = 0.3$. In~\figref{fig:sum}(j), we saw that the system behaves like a semimetal at the lowest temperature. The dispersion of the low-energy quasiparticle peaks is close to that of the bare Hamiltonian $\hat{H}_0$. This is in stark contrast with the K-IVC state at $\kappa = 0.75$, where the dispersion is close to that of the interaction-only Hamiltonian [\figref{fig:sum}(d)]. 
 
 To understand this behavior better, we focus on $\bk=\KM_M$ [\figref{fig:SpecCut}(c)]. At $T=116\,\rm{K}$, we observe a broad maximum around $\omega=0$. 
At lower temperatures, two distinct narrow peaks appear at $\omega \sim \pm 0.7\meV$. These peaks can be identified with the quasi-particle excitation of the Dirac semi-metal. This finite gap originats from the finite-size charging energy $V_{\bq=0}$, which is $0.7 \meV$ for $L=9$, and vanishes in the thermodynamic limit.

Additionally, there are two broad peaks that give rise to the `shoulders' at $\omega \sim 15 \meV$, close to the excitation energy of interaction-only Hamiltonian, shown by the thin dashed lines in \figref{fig:SpecCut}(c) [see also \figref{fig:sum}(j)]. Thus, the low-temperature spectrum at $\KM_M$ contains two distinct features, one near the excitation energy of $\hat{H}_0$ and another at the interaction energy scale. This behavior is reminiscent of that of a Hubbard model, where coherent low-energy quasiparticle peaks can coexist with high-energy `Hubbard bands' at the scale of the local interaction $U$. The similarity suggests that electron correlations play an important role in the semi-metal phase at $\kappa=0.3$. Fig.~\ref{fig:SpecCut}(d) shows the spectral function at $\bk=\Gamma$, which is qualitatively similar to that at $\kappa=0.75$. 

\PRX{Next, we present two additional analyses to characterize further the correlated semi-metal state: (a) the spectral weight at low energies for the Dirac point and (b) averaged double occupancy per spin, valley, and momentum. The former, $\rho_{\mathrm{Dirac}}\equiv(2\pi)^{-1}\int_{-\Delta}^{+\Delta}d\omega A(\omega,K_M)$, is a proxy for the quasiparticle weight $Z$ of the Dirac fermion at the Fermi level, where $\Delta$ is set to twice the charging energy. This quantity converges to the quasiparticle weight in thermodynamic limit. In \figref{fig:semimetal}(a), we find that $\rho_{\mathrm{Dirac}}$ increases with decreasing temperature due to the suppression of thermal spectral broadening and approaches $\rho_{\mathrm{Dirac}}\sim 2.75$ for $L=3$ and $\rho_{\mathrm{Dirac}}\sim 2$ for $L=6,9$ at low temperature. Note that $\rho_{\mathrm{Dirac}} = 4$ for the Slater state of a non-interacting symmetric Dirac semi-metal. 
Hence, roughly half of the spectral weight has been transferred from the Dirac quasi-particles to the continuum. 
This phenomena resembles the reduction of quasiparticle weight in Hubbard model with large interaction parameter in a metallic phase before it undergoes a transition to Mott insulator \cite{Georges2004}.}

\PRX{ \figref{fig:semimetal}(b) shows the averaged flavor-resolved double occupancy, $D \equiv \sum_i \frac{1}{4L^2}\expval{(n_{i,\rm{val}} + n_{i,\rm{con}} - 1)^2}$,
where $i = (s_z,\tau_z,\bk)$ is a collective index for spin, valley, and momentum, and $n_{i,\rm{con}}$ ($n_{i,\rm{val}}$) are the occupation numbers of conduction (valence) band at state $i$, respectively. In the correlated semi-metal state, we observe that $D$ converges to the value around $0.2$ at low temperature with increasing system size. Note that in the ground state of a non-interacting semi-metal, $D\rightarrow 0$ as $L\rightarrow \infty$.
The double occupancy $D$ is a direct measure of fluctuations in $n_i=n_{i,\textrm{val}} + n_{i,\textrm{con}}$. Finite values of $D$ can have multiple origins, e.g., fluctuations between different valleys, spins, or momenta, even in the absence of long-range order. However, fluctuations between conduction and valence band, relevant for the QH and VH channel, do not contribute to $D$.
\figref{fig:semimetal}(b) indicates an extensive double occupancy even in the limit $T\rightarrow 0$ and thus cannot be explained by the usual excitation of the semi-metal in a small region around the Dirac points.}

\PRX{\figref{fig:QMC_channel}(e) indicates the absence of any long-range correlations at low temperature in the semi-metal. There is an enhancement of short-ranged correlations for K-IVC, VH and QH channels at low temperature, while VP and SP channels are highly suppressed. This indicates that the correlated semi-metal state consists of significant quantum fluctuations between conduction and valence bands (VH/QH channels) as well as a quantum fluctuations between different valleys (K-IVC channel).   
}

To summarize the results for $\kappa=0.3$, we find spectra which show features of the semi-metal, but also a considerable transfer of spectral weight to a high-energy correlated mode.
The observed semi-metal phase at $\kappa = 0.3$ is in stark contrast with results of self-consistent HF, which predicts a K-IVC groundstate with a gapped single-particle spectrum with a gap of $16.7 \meV$ at the $\KM_M$ point. The HF calculation with $L=6$ predicts an energy density $E = -3.861 \meV$ per moir\'e unit cell, while the DQMC result at $T=1.8 \TK$ gives $E=-5.388 \pm 0.002 \meV$ per moir\'e unit cell. The energy of the semimetal state is significantly lower than the energy of the HF K-IVC state, which explains why the semi-metal state is preferred over the K-IVC state at $\kappa = 0.3$, and indicates that the ground state wavefunction of the correlated semi-metal is far from a Slater determinant state.

\section{Conclusions and Outlook}
\label{sec:outlook}

In this work, we have demonstrated that the rich physics of magic-angle twisted bilayer graphene at charge neutrality can be simulated using the unbiased, numerically-exact DQMC method, free of the fermion sign problem. This is a rare example where a nearly-realistic model of a strongly-correlated electron system can be fully solved. Our simulations reveal that the nature of the ground state depends sensitively on the bare dispersion bandwidth. Varying the bandwidth, we found either a semi-metal or a K-IVC ordered state. The K-IVC state exhibits a non-trivial evolution with temperature, including a `pseudogap' feature in single-particle density of states that onsets far above the temperature at which the correlation length of the K-IVC order begins to grow rapidly. 
The semi-metal phase appears in a regime where Hartree-Fock would predict an insulating state with a large gap, and exhibits signatures of strong correlations, such as coexistence of low-energy coherent quasiparticles and high-energy correlation-induced modes.

Our work paves the way for more detailed studies  of TBG and direct comparisons to experiments. The phase diagram as a function of the single-particle bandwidth (that can be tuned, e.g., by the distance to the gates) and temperature can be mapped, including any possible intermediate phases between the semi-metal and the K-IVC. The 
finite-temperature crossover regime above the K-IVC can be characterized in more detail. This regime likely hosts strong  bosonic fluctuations, whose entropy (directly calculable in DQMC) can give rise to a Pomeranchuk-like effect \cite{tbg_pomeranchuk, tbg_pomeranchuk2}.  Our method to eliminate the sign problem at small twist angles can be extended to include various physical perturbations, such as strain, a sub-lattice potential, and perpendicular electric and magnetic fields. Furthermore, novel electrical transport at elevated temperatures has been experimentally reported \cite{Cao_2020, Polshyn_2019}, and our setup will help clarify  the intrinsic conductivity of TBG, free of phonons and other non-electronic scattering mechanisms. These investigations will be reported in a forthcoming series of works.

Our method should also be useful to probe strong coupling  superconductivity that leaves an imprint on excitations in the parent insulator. Signatures of incipient superconductivity in future calculations could provide strong support for a purely electronic pairing mechanism.  One such proposal is Skyrmion superconductivity~\cite{KhalafSC,chatterjee2020skyrmion, Christos_2020}. 
Probing for such charge--$2e$ excitations above charge neutrality may shed light on this unusual mechanism for superconductivity.

\acknowledgements

We thank Michael Zaletel, Zhenjiu Wang, and Fakher F. Assaad for explanation of their inspiring work \cite{LL_SO5_QMC}. AV was supported by a Simons Investigator award and by the Simons Collaboration on Ultra-Quantum Matter, which is a grant from the Simons Foundation (651440,  AV). EK  was supported by the German National Academy of Sciences
Leopoldina through grant LPDS 2018-02 Leopoldina fellowship. 
EB and JH were supported by the European Research Council (ERC) under grant HQMAT (Grant Agreement No. 817799), the Israel-US Binational Science Foundation (BSF), CRC 183 of the Deutsche Forschungsgemeinschaft, and by a Research grant from Irving and Cherna Moskowitz. JYL was supported by Gordon and Betty Moore Foundation through Grant GBMF8690 to UCSB and by the National Science Foundation under Grant No. NSF PHY-1748958.
The auxillary field QMC simulations were carried out using the ALF package available at \url{https://alf.physik.uni-wuerzburg.de}.

{\it Note Added:} After the completion of this work, Ref.~\cite{tbgQMC_Meng} appeared which notes absence of a sign problem in certain limits of TBG.

\newpage

\onecolumngrid 

\newpage 

\appendix

\section{Sign Problem}
\label{app:A}

In the Appendix, we revisit the sign-problem in DQMC. We clarify some subtleties that can arise from two different perspectives to look into a sign problems: first quantized and second-quantized Hamiltonians.

In Hirsch's original paper \cite{HirschQMC}, the sign problem of fermion quantum Monte Carlo method was discussed in many-body basis. He showed that in the half-filled Hubbard model on a bipartite lattice, the partition function weight of a given auxiliary field configuration is positive by applying a many-body basis transformation that leaves the trace invariant. However, in the same paper, he also showed how the trace of a  many-body Hamiltonian can be simplified, i.e., one can calculate the trace as the determinant of the corresponding first-quantized Hamiltonian. This form was later exploited by Wu et al. \cite{Zhang2005_sign}, where it was shown that the anti-unitary symmetry ${\cal T}$ of a \emph{single-particle} (or first-quantized) quadratic Hamiltonian obtained by Hubbard-Stratonovich decoupling can guarantee the absence of the sign problem if ${\cal T}^2 = -1$. This argument was further extended by Li et al. \cite{yaoQMC1, yaoQMC2, yaoQMC3}, where they discuss how the set of anti-unitary symmetries of a first quantized (HS-decoupled) Hamiltonain in the Majorana representation can guarantee the absence of sign problem. 
The Majorana representation is nice, since the unitary transformation of a first-quantized Hamiltonian in the electron representation cannot accommodate particle-hole symmetry, which is a valid basis transformation in a many-body level. The problem is that single-particle Hamiltonian encodes the information about the many-body physics with an explicit $U(1)$ number conservation symmetry. Therefore, if we want to discuss a generic particle number changing transformation at a first-quantized level, it advantageous to use the Majorana representation.

However, without bothering to reformulate the problem in Majorana basis, we can simply work with a (HS-decoupled) many-body Hamiltonian as Hirsch originally discussed. There is a subtlety. In his original discussion, he only considered a bipartite system with a real Hamiltonian and associated many-body particle-hole transformation $\cC$, and it was not important whether $\cC$ is unitary or anti-unitary at many-body level. For a more general approach, we have to understand the role of anti-unitary many-body symmetries, which will be provided below. First note that $(\bra{n} M \ket{m})^* = \bra{m} M^\dagger \ket{n}$. The following identity is useful in the discussion of anti-unitary transformation:
\begin{equation} \label{eq:anti-unitary}
    A = \Tr M \equiv \sum_n \bra{n} M \ket{n}, \quad \Rightarrow \quad A^* = \sum_n (\bra{n} M \ket{n})^* = \sum_n \bra{n} M^\dagger \ket{n} = \Tr M^\dagger = \Tr M^*,
\end{equation}
where the last equality comes from the invariance of the trace under transpose. This part is important as the trotterized partition function cares about the order of matrix products. Let ${\cal K}$ be the complex-conjugation. Now define a general anti-unitary symmetry ${\cal T} = {\cal U}{\cal K}$, where ${\cal U}$ is a many-body unitary basis transformation. Note that if ${\cal U}$ is a many-body \emph{unitary} basis transformation, it can be shown that
\begin{align} \label{eq:sym_transform}
    \Tr_c e^{-\beta \hat{H}} = \Tr_{c} {\cal U}^{-1} e^{-\beta {\cal U}\hat{H}{\cal U}^{-1}} {\cal U} = \Tr_{{\cal U}c} e^{-\beta {\cal U}\hat{H}{\cal U}^{-1}} = \Tr_{c} e^{-\beta {\cal U}\hat{H}{\cal U}^{-1}}
\end{align}
where the last equality comes from the cyclic property of trace. 
By combining \eqnref{eq:sym_transform} and \eqnref{eq:anti-unitary}, we can show that
\begin{equation} \label{eq:anti-unitary_theorem_app}
    A \equiv \Tr_c \prod_l e^{- \hat{H}_l} \quad \Rightarrow \quad A^* = \Tr_c \prod_l e^{- {\cal T} \hat{H}_l {\cal T}^{-1}}.
\end{equation}

To illustrate the point, consider a bipartite Hubbard model with nearest-neighbor hopping terms and on-site repulsive interaction at charge-neutrality:
\begin{equation}
    \hat{H} = \sum_{\expval{ij}} \sum_s c^\dagger_{i,s} H_{ij} c_{j,s}^{\,} + \frac{U}{2} \sum_i  (n_{i,\uparrow} + n_{i,\downarrow}-1)^2
\end{equation}
The Hirsch's original approach performed a HS-decoupling into a spin-channel when the interaction is repulsive ($U>0$) so that the resulting decoupled interaction term is real. However, even when $U>0$, we can also decouple the interaction into a density channel with a price that the interaction term would have an imaginary coefficient, thus non-Hermitian. Under Hubbard-Stratonovich transformation, the trotterized on-site repulsive interaction term would be given as the following:
\begin{align}
    e^{ -\frac{U \delta}{2} (n_\uparrow  + n_\downarrow - 1)^2 }  &= \frac{1}{2} \sum_{\eta = \pm 1} e^{i \alpha \eta (n_\uparrow + n_\downarrow - 1)}  = \frac{1}{2} \sum_{\eta = \pm 1} e^{i \alpha \eta(n_\uparrow -1/2)} \cdot  e^{i\alpha \eta( n_\downarrow - 1/2)}
\end{align}
where $\alpha = \cos^{-1} e^{-U \delta /2}$ ($U>0$). Under density-channel decoupling, both spins have the same HS-decoupled Hamiltonians, implying that $W_\uparrow[\eta(\tau)] = W_\downarrow[\eta(\tau)]$ (c.f. \eqnref{eq:Z}). Now, consider a many-body basis transformation ${\cal U}$:
\begin{equation}
    {\cal U}: c^\dagger_{r,s} \mapsto e^{i \vec{K} \cdot \vec{r}} c^{\,}_{r,s}, \qquad \vec{K}=(\pi,\pi)
\end{equation}
which exchanges particles and holes. Note that ${\cal U}$ alone is not a symmetry yet. If we consider an anti-unitary operator ${\cal C} \equiv {\cal U} {\cal K}$, then we see that
\begin{align}
    {\cal C} \Big[ \sum_{\expval{ij}} \sum_s c^\dagger_{i,s} H_{ij} c_{j,s}^{\,} \Big]  {\cal C}^{-1} & = -\sum_{\expval{ij}} \sum_s c^{\,}_{i,s} H^*_{ij} c_{j,s}^{\dagger} = \sum_{\expval{ji}} \sum_s c^\dagger_{j,s} H_{ji} c_{i,s}^{\,} \nonumber \\
    {\cal C} \qty[ i \alpha s_{r,\tau} \qty(\hat{n}_{r,\eta} - \frac{1}{2})]  {\cal C}^{-1} & = (-i) \alpha \eta_{r,\tau}  \qty(\frac{1}{2} - \hat{n}_{r,\eta}) = i  \alpha \eta_{r,\tau} \qty( \hat{n}_{r,\eta}- \frac{1}{2}).
\end{align}
Therefore, \eqnref{eq:anti-unitary_theorem_app} implies the following:
\begin{equation}
    W^*_\uparrow[\eta(\tau)] = W_\uparrow[\eta(\tau)], \quad W^*_\downarrow[\eta(\tau)] = W_\downarrow[\eta(\tau)],
\end{equation}
where $W$ is the weight for a given auxiliary field configuration (c.f. \eqnref{eq:W})
Since $W_\uparrow[\eta(\tau)] = W_\downarrow[\eta(\tau)]$, we conclude that
\begin{equation}
        W = \Tr_{\eta(\tau_l)}  \Tr_{\eta(\tau_l)}  \abs{ W_\uparrow[\eta(\tau)]  }^2 > 0.
\end{equation}
This is particularly nice since spin-channel decoupling is not possible for systems with higher spins, i.e., more than two flavors of electrons per site. Even when we have two flavors, this is often a better scheme as the density-decoupling respects  the spin-rotation symmetry, while the $S_z$-decoupling does not.

\section{Twisted Bilayer Graphene and Symmetry} 
\label{app:B}

\begin{figure}[t]
    \centering
    \includegraphics[width = 0.75 \textwidth]{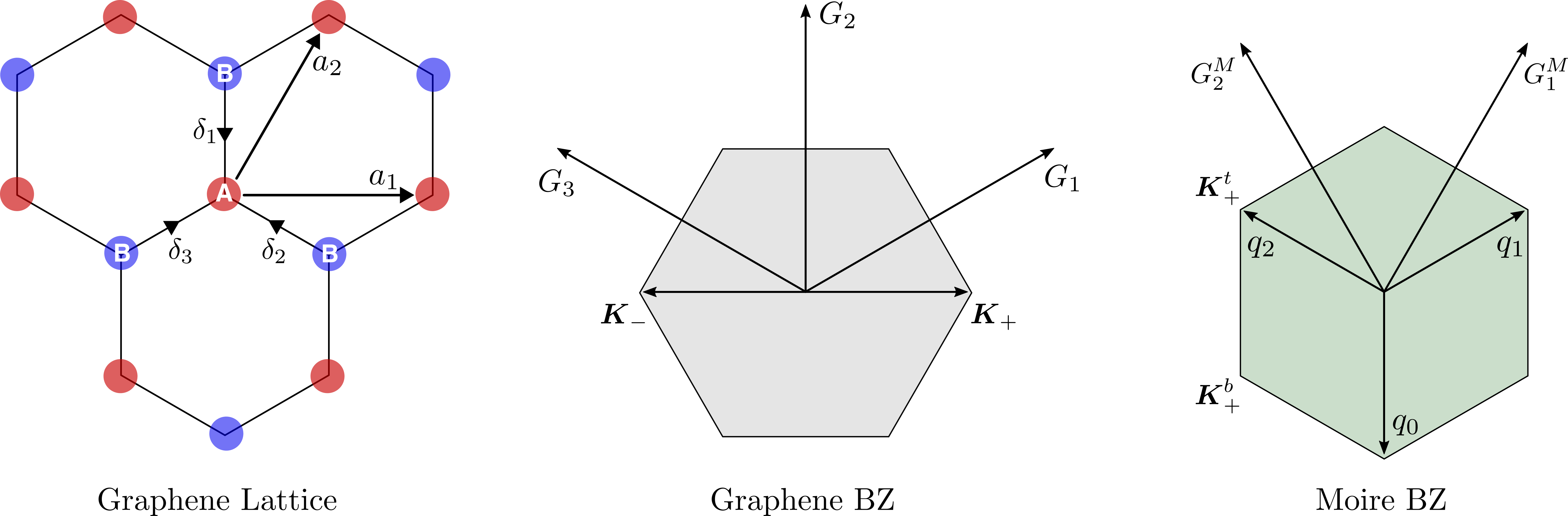}
    \vspace{0.05in}
    \caption{Real space lattice and Brillouin zone for a single graphene layer, and Brillouin zone origniated from $\bK_+$ valley for a twisted bilayer graphene. The figure illustrates all vectors labeled for the discussion in the manuscript.}
\end{figure}

Let $\theta$ be a (counterclockwise) twist angle of top layer with respect to the bottom layer. Then, the TBG hamiltonian for valley $\tau$ has the following form:

\begin{align}\label{eq:HABAB}
    H_\tau &= \sum_{\bk} \Big[\,  c_{\bk,\tau,t}^\dagger h_{\theta/2, \tau }(\bk) c^{}_{\bk,\tau,t}  +  c_{\bk,\tau,b}^\dagger h_{-\theta/2, \tau }(\bk) c^{}_{\bk,\tau,b} \nonumber \\
    & \quad + \sum_n \qty( c^\dagger_{\bk,\tau,b} T_{n,\tau} c^{~}_{\bk+\bq_n,\tau,t} +  c^\dagger_{\bk+\bq_n,\tau,t} T^\dagger_{n,\tau} c^{~}_{\bk,\tau,b} ) \Big], 
\end{align}
where $t,b$ denotes top and bottom layers and $h_{\theta,\tau}(\bk)$ is the matrix for the Dirac Hamiltonian rotated by $\theta$ in counter-clockwise direction. In other words, $h_{\theta,\tau}(\bk) = h_{0,\tau}(R_{-\theta}\bk)$, and the moir\'e interlayer hopping term is defined as $T_{n,\tau}  =  w_0 + w_1  e^{2\pi i n \sigma_z \tau  /3} \sigma_x e^{-2\pi i n \sigma_z \tau  /3}$. 
Here, $c^\dagger_{\bk,\tau,t/b}$ is a 4-components (spin and sublattice) electron creation operator with momentum $\bk$ measured with respect to individual Dirac points at valley $\tau$ for top/bottom layers, respectively. In this paper, we choose the convention where Dirac Hamiltonian is given as $h_{0,\tau}(\bk) \equiv \tau k_1 \sigma_1 + k_2 \sigma_2 = \bk_\tau \cdot \bsigma $ where $\bk_\tau \equiv (\tau k_1, k_2)$. Now, if we rotate momentum by angle $\varphi$ in counter-clockwise direction, we get the following:
\begin{align}
    (R_\varphi \bk)_\tau \cdot \bsigma &= \tau (k_1 \cos \varphi - k_2 \sin \varphi) \sigma_1  + (k_1 \sin \varphi + k_2 \cos \varphi) \sigma_2 \nonumber \\
    & = \tau k_1 ( \cos \varphi  \sigma_1 + \sin \varphi \sigma_2 \tau ) + k_2 (- \sin \varphi \tau \sigma_1  + \cos \varphi \sigma_2).
\end{align}
Using the fact that $e^{i \varphi \sigma_3 \tau} = \cos \varphi + i \sin \varphi \sigma_3 \tau$, we can show that
\begin{align}
    e^{-i \varphi \sigma_3 \tau/2} \tau \sigma_1 e^{i \varphi \sigma_3 \tau/2} &= \tau \sigma_1 e^{i \varphi \sigma_3 \tau}  = \tau (\cos \varphi  \sigma_1 + \sin \varphi \sigma_2 \tau)  \nonumber \\
    e^{-i \varphi \sigma_3 \tau/2} \sigma_2 e^{i \varphi \sigma_3 \tau/2} &= \sigma_2 e^{i \varphi \sigma_3 \tau}  = \cos \varphi \sigma_2 - \sin \varphi \tau \sigma_1 .
\end{align}
Therefore, $(R_\varphi \bk)_\tau \cdot \bsigma = (\bk_\tau \cdot \bsigma) e^{i \varphi \sigma_3 \tau}$. As $\varphi = - \theta \mu_z/2$, we find that $(R_\varphi \bk)_\tau \cdot \bsigma = (\bk_\tau \cdot \bsigma) e^{- i \theta \sigma_3 \tau \mu_3/2}$. Therefore, we can unify the descriptions for different layers and valleys in the following form:
\begin{align}  
    \hat{H} &= \sum_{\bk}  c^\dagger_\bk H(\bk) c_\bk + \sum_{\bk, \bk'} c^\dagger_{\bk'} T(\bk',\bk)  c^{\,}_\bk   \nonumber \\
    H(\bk) &= \hbar v_F \qty( k_x \sigma_x \tau_z + k_y \sigma_y ) e^{-i \theta \sigma_z \mu_z \tau_z /2} 
    \nonumber  \\
    T(\bk',\bk) &= \frac{1}{2} \sum_n \delta_{\bk', \bk - \tau_z \bq_n}   \qty( \mu_x - i \mu_y ) T_n  +  \frac{1}{2} \sum_n \delta_{\bk', \bk+ \tau_z \bq_n}  T^\dagger_n \qty( \mu_x + i \mu_y )   
\end{align}
where $\sigma$, $\tau$ and $\mu$ act on sublattice, valley and layer respectively.  Here $\bq_n = {\cal R}_{2\pi n/3}\, \bq_0$, where ${\cal R}_{\varphi} = e^{i \varphi \sigma_y}$ is a rotation matrix, $\bq_0 = \frac{8\pi \sin( \theta/2)}{3 \sqrt{3} a_0} (0,-1)^T$, and $a_0 = 1.42 \textrm{ \AA}$ is the intra-layer distance between carbon atoms. Also, note that $T_n^\dagger = T_n$. $c^\dagger_\bk$ is a 16-components creation operator, whose flavors contains spin, valley, layer, and sublattice.

\begin{figure}[t]
    \centering
    \includegraphics[width = 0.7\textwidth]{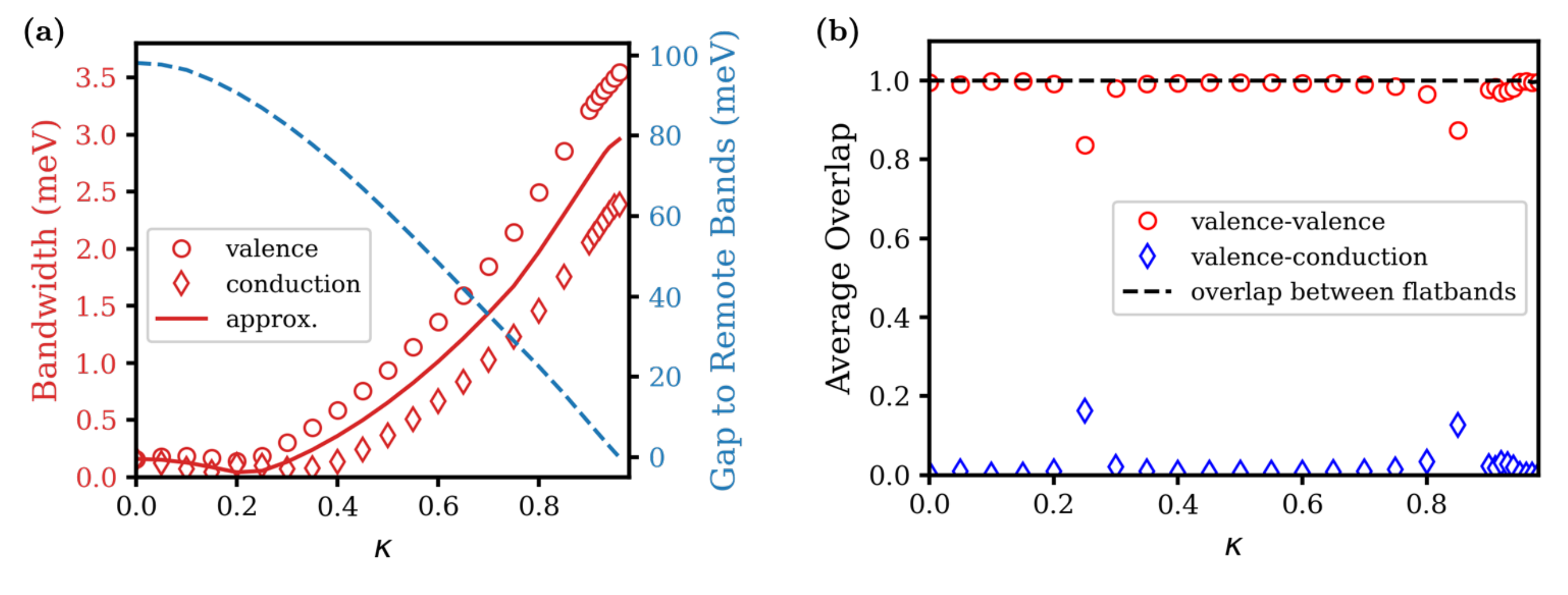}
    \caption{\label{fig:overlap} Here we plot comparisons between the original BM Hamiltonian and its small-angle approximated Hamiltonian  with varying $\kappa = w_0/w_1$ at $\theta=1.05$, $v_0 = 2700 \meV$, and $w_1 = 105 \meV$. (a) The discrepancy of bandwidth between the original TBG model and approximated chiral TBG model. Here we also plot the gap between the conduction band and second conduction (remote) band. Comparing left and right y-axes, we notice that the flat bands are well-separated from remote bands in the wide range of $\kappa$. (b) Wavefunction overlaps between the original BM model and small-angle approximated BM model for valence bands, valence/conduction bands, and flatband subspaces. As a whole, flat-band subspaces between two Hamiltonians are almost identical until the gap to remote bands close around $\kappa = 1$.  }
\end{figure}

To examine the symmetry properties more closely, we apply the following basis transformation:
\begin{equation}\label{eq:basis_transform}
    U_g = e^{-i \theta \sigma_z \mu_z \tau_z/4}, \qquad c_\bk \mapsto U_g c_\bk, \qquad H(\bk) \mapsto U_g H(\bk) U_g^\dagger = \hbar v_F \qty( k_x \sigma_x \tau_z + k_y \sigma_y ) .
\end{equation}
Under this transformation, we can remove the explicitly $\theta$-dependent phase factor in $H(\bk)$, while $T(\bk',\bk)$ is slightly modified, resulting in
\begin{align} \label{eq:tbg_ham_supp}
    \hat{H} &= \sum_{\bk}  c^\dagger_\bk H(\bk) c_\bk + \sum_{\bk, \bk'} c^\dagger_{\bk'} T(\bk',\bk)  c^{\,}_\bk   \nonumber \\
    H(\bk) &= \hbar v_F \qty( k_x \sigma_x \tau_z + k_y \sigma_y )
    \nonumber  \\
    T(\bk',\bk) &= \frac{1}{2} \sum_n \delta_{\bk', \bk - \tau_z \bq_n}   \qty( \mu_x - i \mu_y ) \tilde{T}_n  +  \textrm{h.c.} 
\end{align}
where
\begin{equation}
\tilde{T}_n  =  w_0 e^{-i \theta \sigma_z \mu_z \tau_z/2} + w_1  e^{2\pi i n \sigma_z \tau_z /3} \sigma_x e^{-2\pi i n \sigma_z \tau_z /3}.   
\end{equation}
Now, consider a many-body basis transformation
\begin{align}
    {\cal C} : \quad c^\dagger_\bk &\mapsto \tau_x \sigma_x \mu_y  c^{\,}_\bk \nonumber \\
     i &\mapsto -i 
     \label{eq:C}
\end{align}
which is a proper symmetry when $w_0=0$. The operator $\cC$ maps $w_0 e^{-i \theta \sigma_z \mu_z \tau_z/2}$ to $w_0 e^{i \theta \sigma_z \mu_z \tau_z/2}$. For a small $\theta$, the discrepancy is indeed very small. If we ignore $\theta$-dependence in the transformed Hamiltonian, i.e., setting $\tilde{T}_n$ as
\begin{equation}
\tilde{T}_n  =  w_0 + w_1  e^{2\pi i n \sigma_z \tau_z /3} \sigma_x e^{-2\pi i n \sigma_z \tau_z /3},   
\end{equation}
enhances the approximate symmetrz $\cC$ to an exact one. To test the validity of this approximation, we examined the wavefunction overlap between eigenstates of the actual Hamiltonian and approximated Hamiltonian as function of $w_0$ and $\bk$ in \figref{fig:overlap}. As a whole, the flat-band subspace of the two Hamiltonians are almost identical until the gap to remote bands close around $\kappa = 1$.  Individually, the valence band overlap between two Hamilonians drops around $\kappa \sim 0.25$ and $\kappa \sim 0.85$. The discrepancy at $\kappa \sim 0.25$ is physically not very important since the bandwidth illustrated in (a) is extremely flat, implying that valence and conduction bands are almost degenerate. Therefore, compared to the interaction energy scale, this discrepancy can be ignored. The discrepancy at $\kappa \sim 0.85$ might have a physical significance, since this implies that the valence and conduction bands hybridize at  certain $\bk$-points other than at $\KM$ and $\KM'$. Therefore, we confirm that the approximation works well for $w_0 < 0.8$.

\section{Hartree-Fock calculation of the bare dispersion} \label{app:D}

In this section, we review the Hartree-Fock calculation, which is used to calculate the bare dispersion $\hat{H}_0$ 
In the BM model Hamiltonian, we obtain the single-particle dispersion under the assumption that the interaction does not affect the single-particle physics. However, note that its tight-binding parameters are extracted from first principle calculations and scattering experiments, where interaction effects are taken into account. Therefore, the obtained BM model dispersion can be considered as an effective bandwidth resulting from a bare quadratic terms $\hat{H}_0$ and the Coulomb interaction $\hat{V}$ \cite{Liu2021, BultnickKhalaf2020}. In other words,
\begin{equation}
    \hat{H}_\textrm{BM} = \qty[ \hat{H}_0 + \hat{V}\, ]_{\Psi}, \qquad \hat{V} = \frac{1}{2} \sum_\bq \hat{\rho}_\bq V_\bq  \hat{\rho}_{-\bq}
\end{equation}
for some ground-state wavefunction $\Psi$. If we want to simulate the full-interacting Hamiltonian, we should use the bare dispersion $\hat{H}_0$ instead of naively using $\hat{H}_\textrm{BM}$, to avoid double-counting. By choosing a proper reference state, one can calculate the bare dispersion as
\begin{equation} \label{eq:bareH_app}
    \hat{H}_0 = \hat{H}_\textrm{BM} - \big[ \hat{V}\, \big]_{\Psi}.
\end{equation}
When $\Psi$ is a Slater state, its electronic correlation is fully captured by a single-particle density matrix $P_0$ and we can calculate the mean-field contribution of the interaction $\hat{V}$ with respect to $P_0$ using the Hartree-Fock method. In the following, we elaborate on the details of the Hartree-Fock calculation and how the bare dispersion $\hat{H}_0$ and interaction-only spectrum $[\hat{V}]$ can be obtained.

\begin{figure}[t]
    \centering
    \includegraphics[width = 0.8 \textwidth]{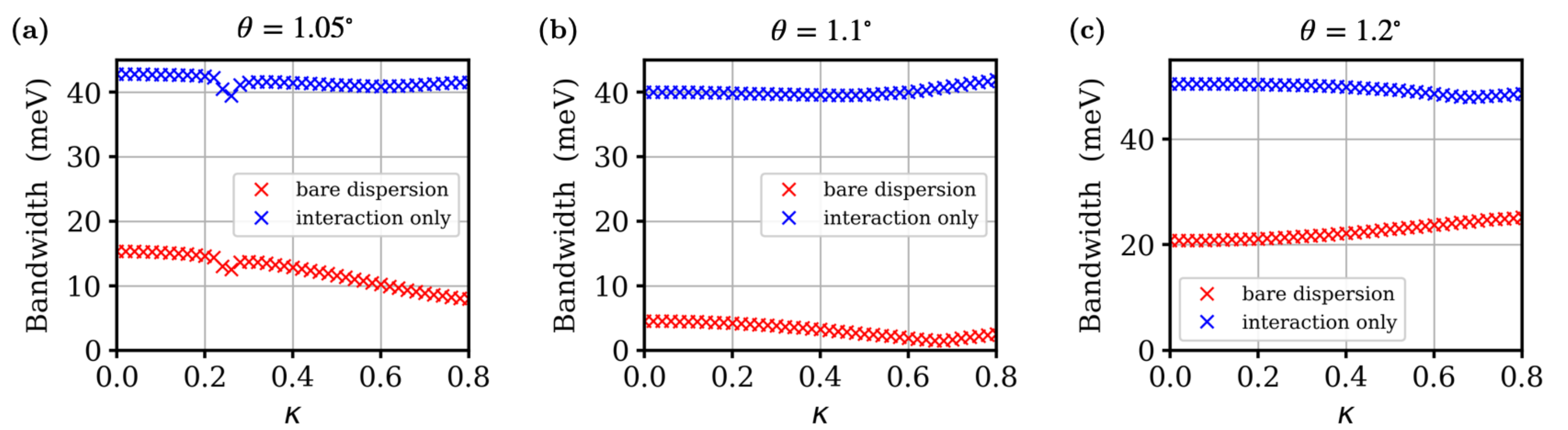}
    \caption{\label{fig:different_angle} The bandwidths of electron-hole excitation spectra for bare dispersion $\hat{H}_0$ and the interaction-only model at (a) $\theta=1.05^\circ$, (b) $\theta=1.1^\circ$, and (c) $\theta=1.2^\circ$ under the subtraction scheme in \eqnref{eq:bare_mod_density}. Here we use $\frac{\hbar v_f}{3a_0/2} = 2700 \meV$, $w_0 = 105 \meV$, and $6 \times 6$ system size. The result shows that the bare dispersion near the BM-model magic angle $\theta = 1.05^\circ$ is not minimal. Rather, the bare dispersion is  suppressed at a slightly higher angle of $1.1^\circ$, while at $1.2^\circ$, it increases back. Therefore, the so-called `magic angle's for the BM dispersion and bare dispersion are somewhat different. Around the twist angle $1.2^\circ$, the bare dispersion is significantly larger than that of $\theta=1.05^\circ$ throughout the whole range of $\kappa$. Based on this figure, we infer that the ground-state at this larger angle would be semi-metallic even at $\kappa=0.75$. }
\end{figure}

Let $P_0$ be the one-body projector for the slater reference state $\Psi_{0}$. Following the standard Hartree-Fock calculations and projection to the flat bands \cite{Repellin2020}, the bare Hamiltonian should be given as the following:
\begin{align} \label{eq:bare_density}
    \big[ \hat{V} \big]_{P_0} &= \sum_\bk \sum_{mn \in {\cal I} } c^\dagger_{\bk, m} \Big(\, \xi^H_{\bk, mn}[P_0] + \xi^F_{\bk, mn}[P_0]  + \xi^\textrm{Norm}_{\bk,mn} - \xi^H_{\bk, mn}[\Delta] - \xi^F_{\bk, mn}[\Delta] \Big) c^{\,}_{\bk, n} \nonumber \\ \hat{H}_0 &= \hat{H}_\textrm{BM}  - \sum_\bk \sum_{mn \in {\cal I} } c^\dagger_{\bk, m} \Big(\, \xi^H_{\bk, mn}[P_0-\Delta] + \xi^F_{\bk, mn}[P_0-\Delta]  + \xi^\textrm{Norm}_{\bk,mn}  \Big) c^{\,}_{\bk, n} 
\end{align}
where $\cal I$ is the set of indices of the flat-bands. The second equality comes from the fact that Hartree and Fock terms are linear in the projector matrix, as in \eqnref{eq:HFterms}. Here, the first two terms are from the HF contributions of the reference Slater state represented by $P_0$,  and the third term is from the normal-ordering. Note that, for the Hartree-Fock calculation, one should first normal order the term $\frac{1}{2} \sum_\bq V_\bq \rho_\bq \rho_{-\bq} = \frac{1}{2} \sum_\bq V_\bq :\rho_\bq \rho_{-\bq}: + \xi^\textrm{Norm}$. The last two terms stem from the projected-out occupied bands, whose one-body projector is defined as
\begin{equation}
    \Delta_{ij} = \begin{cases} \delta_{ij} & \quad \textrm{if } i,j \in \bar{\cal I}_o \\ 0 & \quad \textrm{otherwise}  \end{cases}
\end{equation} 
where $\bar{\cal I}$ is the set of indices of the projected-out remote bands, and the subscript $o$ (or $u$) are for occupied (or unoccupied) bands. Each contribution is defined as the following:
\begin{align} \label{eq:HFterms}
    \xi^H_{\bk,mn}[M] &= \sum_{\bG} [\Lambda^\dagger(\bk,\bG)]_{mn} \sum_{\bk'} \tr(M^T(\bk') \cdot \Lambda(\bk',\bG)) \nonumber\\
    \xi^F_{\bk,mn}[M] &= -\sum_\bq V_\bq \qty[ \Lambda^\dagger(\bk,\bq) M^T(\bk+\bq) \Lambda(\bk,\bq) ]_{mn} \nonumber\\
    \xi^\textrm{Norm}_{\bk,mn} &= \frac{1}{2} \sum_\bq V_{\bq} \qty[ \Lambda^\dagger(\bk,\bq) \Lambda(\bk,\bq) ]_{mn}  = - \frac{1}{2} \xi^F_{\bk,mn}[\mathbb{I}_{\cal I}]
\end{align}
where we used the fact that $V_\bq = V_{-\bq}$, and the indices for $\Lambda$ run over ${\cal I} \cup \bar{\cal I}$. Here, $\mathbb{I}_{\cal I}$ is the matrix which is the identity on the flat-band block and zero on the other.

As discussed in \cite{Repellin2020}, the form of the bare Hamiltonian depends on what forms of interaction we use. In \eqnref{eq:bare_density}, the interaction is given as the $\sim \rho_\bq \rho_{-\bq}$, which is the raw density-density form, where the density is measured from the empty filling, not the charge neutrality. For our purposes, let us use the density $\bar{\rho}$ which is measured with respect to the charge neutrality and choose the interaction accordingly, i.e., $\hat{V} \sim (\rho_\bq - \rho^0_\bq)(\rho_{-\bq} - \rho^0_{-\bq}) = \bar{\rho}_\bq \bar{\rho}_{-\bq}$. With this modification, the bare Hamiltonian is calculated as the following:
\begin{align} \label{eq:bare_dev_density}
    \hat{H}_0 &= \hat{H}_\textrm{BM}  - \sum_\bk \sum_{mn \in {\cal I} } c^\dagger_{\bk, m} \Big(\, \xi^H_{\bk, mn}[P_0-\Delta] + \xi^F_{\bk, mn}[P_0-\Delta]  + \xi^\textrm{Norm}_{\bk,mn} - \xi^\textrm{back}_{\bk,mn} \Big) c^{\,}_{\bk, n} \nonumber \\
    &= \hat{H}_\textrm{BM}  - \sum_\bk \sum_{mn \in {\cal I} } c^\dagger_{\bk, m} \qty(\, \xi^H_{\bk, mn}\qty[P_0-\Delta-\frac{1}{2} \mathbb{I}_{\cal I}] + \xi^F_{\bk, mn}\qty[P_0-\Delta-\frac{1}{2}\mathbb{I}_{\cal I}]  ) c^{\,}_{\bk, n} \nonumber \\
    \xi^\textrm{back}&=\frac{1}{2} \sum_\bq V_\bq \qty(  \rho_{\bq} \rho^0_{-\bq} +  \rho^0_{\bq} \rho_{-\bq} ) = \frac{1}{2} 
    \xi^H[\mathbb{I}_{\cal I}]
\end{align}
Interestingly, quadratic terms originating from the normal ordering and background charge density can be incorporated into the Hartree and Fock terms as above. Furthermore, in the case where $P_0-\Delta = \frac{1}{2}(\bI + Q)$ is 
particle-hole symmetric, i.e., for first quantized expression of PH symmetry $\cC$ in \eqnref{eq:C}, $\textrm{(PH)} \, Q\, \textrm{(PH)}^{-1} = -Q$, and where $\textrm{(PH)} \, \Lambda(\bk,\bq)\, \textrm{(PH)}^{-1} = \Lambda(\bk,\bq)$, one can prove that $\xi^H$ contribution vanishes. Since this is the case under small-angle approximation, we are left with only the Fock contribution. Therefore, in this case our total Hamiltonian $\hat{H} = \hat{H}_0 + \hat{V}$ is defined as the following:
\begin{align} \label{eq:bare_mod_density}
    \hat{H}_0 &= \hat{H}_\textrm{BM}  - \sum_\bk \sum_{mn \in {\cal I} }  \xi^F_{\bk, mn}\Big[P_0-\Delta-\mathbb{I}_{\cal I}/2 \Big] \cdot c^\dagger_{\bk, m} c^{\,}_{\bk, n} \nonumber \\
    \hat{V} &= \frac{1}{2} \sum_\bq \bar{\rho}_\bq V_\bq \bar{\rho}_{-\bq} - \frac{1}{2} \sum_\bG V_\bG \rho^0_\bG \rho^0_{-\bG}, \qquad \bar{\rho}_\bq \equiv \rho_\bq - \frac{1}{2} \sum_{\bk, \bG} \delta_{\bq,\bG} \tr_{\cal I} \Lambda(\bk,\bG)
\end{align}
This formulation implies that, if we naively use $\hat{H}_\textrm{BM}$ for the bare dispersion, it is equivalent to take $P_0 - \Delta \equiv \mathbb{I}_{\cal I}/2$. Physically, such a choice corresponds to the case where the reference state is an infinite temperature ensemble.

Now, we are ready to calculate the bare dispersion following the \eqnref{eq:bare_mod_density} with a proper choice of $P_0$. In \cite{Liu2021}, one takes $P_0$ to be the one-body projector for the groundstate of the BM Hamiltonian. In \cite{BultnickKhalaf2020, Soejima2020}, the reference state $\Psi_0$ is taken to be the groundstate of decoupled bilayer graphenes. In the second approach, the underlying assumption is that tight-binding parameters including Dirac velocity are extracted for isolated graphene layer, which will be renormalized under the interlayer coupling. Here we take the second approach, where we only considered the flat band subspace and ignored the contribution from remote bands. The results are illustrated in \figref{fig:different_angle}, where we plot the bandwidth $W$ of $\hat{H}_0$ as function of $\kappa = w_0/w_1$ for three different twist angles $\theta=1.05,\,1.1,\,1.2$ (Red crosses). We also show the excitation spectrum of the strong coupling limit (interaction-only with $\hat{H} = \hat{V}$) at charge neutrality. Since any polarized Slater state listed in \tabref{tab:OP} is the exact groundstate in this limit, we can chose the valley-polarized state and calculated the exact electronic excitation spectrum using the following equation:
\begin{align}  
    \big[\,\hat{V} \,\big]_{P} &=  \sum_\bk \sum_{mn \in {\cal I} }  \xi^F_{\bk, mn}\Big[P-\Delta-\mathbb{I}_{\cal I}/2 \Big] \cdot c^\dagger_{\bk, m} c^{\,}_{\bk, n} 
\end{align}
where we use $P_{ij} = 1$ if $i=j$ are in $+$ valley, and 0 otherwise.

\section{Details of the Numerical Steps} 
\label{app:C}

\subsection{Quantum Monte Carlo Hamiltonian}

In the DQMC, we simulate the following Hamiltonian in momentum space:
\begin{align}\label{eq:ham_QMC_supp}
    \hat{H} & = \hat{H}_0 + \hat{V} \nonumber \\
    \hat{H}_0 &= \sum_{s,\tau} \qty[ \sum_{n,m} H^0_{\tau,\bk,nm} c^\dagger_{\bk,s,\tau,n} c^{\,}_{\bk,s,\tau,m}]  \nonumber \\
    \hat{V} &= \frac{1}{8} \sum_\bq V_\bq \qty[ (\bar{\rho}_\bq + \bar{\rho}_{-\bq})^2 - (\bar{\rho}_\bq - \bar{\rho}_{-\bq})^2 ] 
\end{align}
where $\hat{H}_0$ is defined in \eqnref{eq:bareH_app}, which is block-diagonal in spin and valley flavors. Here the Fourier transformation of the interaction $V_\bq$ is defined as the following for a naive screened Coulomb interaction with screening length $d$:
\begin{equation}
    V_\bq = \frac{1}{A} \int \dd^2 r\, e^{i\bq r} \qty[ \frac{e^2}{4\pi \epsilon_0 \epsilon r} e^{- \abs{r}/d_s}]  =  \frac{V_0 d}{\sqrt{1 + \abs{\bq}^2 d^2}}, \qquad V_0 = \frac{e^2 }{2 \epsilon \epsilon_0 } \frac{1}{A} = \frac{e^2}{\sqrt{3} \epsilon \epsilon_0 a_M^2} \frac{1}{N_M}
\end{equation}
where $A = N_M A_0$ is the total area of the system, $A_0$ is the area of a moir\'e unit cell, and $N_M$ is the number of moir\'e unit cells in the systems. $A_0$ is given as $A_0 = \sqrt{3} a_M^2/2$, where $\abs{a_M} = \sqrt{3} a/2 \sin \frac{\theta}{2} = 13.4\textrm{ nm}$ at $\theta = 1.05^\circ$ with  $a = 1.42 \,\textrm{\AA}$. For realistic model, one can  consider a more accurate description that involves gate screening, but they are almost the same in their functional forms. For the doubly gated case with gate distance $d$, we have the following screened Coulomb interaction in the momentum space:
\begin{equation}
    V_\bq = \frac{V_0 \tanh(\abs{\bq} d)}{\abs{\bq}}, \qquad V_0 = \frac{e^2}{2 \epsilon \epsilon_0 } \frac{1}{A} = \frac{e^2}{\sqrt{3} \epsilon \epsilon_0 a_M^2} \frac{1}{N_M}
\end{equation}
At $\bq = 0$, we obtain the charging energy $V_\textrm{CE}= \frac{1}{2}V_0 d$. This is important as the value would contribute to the spectral function even for a gapless excitations. In the parameter we use, note that $V_\textrm{CE}  \sim \frac{58}{L^2} \, \textrm{meV}$. At $L=9$, the charging energy is 0.7\,meV.

\subsection{Symmetry}
\label{app:sym_analysis}

Once we project onto the flat-band subspace, there are total 8 flavors of electrons: spin $s_z = \pm1$, valley $\tau_z = \pm1$, and band $n_z = \pm 1$. 
Under the absence of dispersion and interaction, the system is invariant under $\U(8)$ symmetry, i.e., for ${\cal U} \in \U(8)$, $\hat{H}$ is invariant under $d_\bk \mapsto {\cal U} d_\bk$. There are 64 generators that are given as $s_\mu \tau_\nu n_\lambda$, with $\mu,\nu,\lambda = 0,x,y,z$. Of course, in realstic system this should reduce into $\U(2)_K \times \U(2)_{K'} \sim \U(1)_\textrm{el} \times \U(1)_\textrm{valley} \times \SU(2)_{K} \times \SU(2)_{K'}$. As we switch the dispersion and interaction back on, the symmetry is reduced, as elaborated in \cite{BultnickKhalaf2020}.

\bigskip

\begin{table}[t]
\setlength\tabcolsep{6pt}
\renewcommand{\arraystretch}{1.25}%
\begin{tabular}{|c|c|c|c|c|}
\hline
Symmetries & Microscopic & Sublattice Basis & Band Basis I & Band Basis II \\
\hline
\hline
${\cal T}$ ($\bk \mapsto -\bk$) & $\tau_x \cK$ & $\tau_x \cK$ & $\tau_x n_z \cK$ & $\tau_x \cK$ \\ 
\hline 
$C_2$ ($\bk \mapsto -\bk$) & $\tau_x \sigma_x$ & $\tau_x \sigma_x e^{i \theta (\bk)} $ & $\tau_x$ & $\tau_x$ \\ 
\hline 
$C_2 \cT$  & $\sigma_x \cK$ & $e^{i \theta (\bk)}  \sigma_x \cK $ & $n_z \cK$ & $\cK$ \\ 
\hline
Small-angle Approx. $\cC$  & $\tau_x \sigma_x \mu_y$ & $\tau_y \sigma_y $ & $\tau_x n_x$  & $\tau_x n_y$ \\ 
\hline
Sublattice $\cC_\textrm{sub}$ (at $\kappa = 0$) & $\sigma_z$ & $ \sigma_z$ & $\tau_z n_x$ & $\tau_z n_y$ \\ 
\hline
\end{tabular}

\caption{\label{tab:sym_rep} Representations of symmetry operators in different basis. Sublattice basis is constructed in \cite{BultnickKhalaf2020}, and the band basis I is explicitly constructed in the discussion. The band basis II can be obtained for a different gauge choice. Among these operators, $\cT$ and $C$ flip the momentum $\bk$. Note that the above choice is not unique. They should satisfy the following relations: $[\cT, C_2 \cT] = [\cT, \cC_\textrm{sub}] = \{ \cT, \cC \} =0 $, $\{C_2 \cT, \cC_\textrm{sub}\} = \{ C_2 \cT, \cC \} =0 $, $\{ \cC, \cC_\textrm{sub} \} = 0$. Furthermore, in the band basis, chiral symmetries $\cC_\textrm{sub}$ and $\cC$ should anticommute with $n_z$. Depending on the detailed setting of the simulation, we may use one basis over the other. Note that $\cC_\textrm{sub}$ does not have a proper representation on the flat-band subspace if we are away from the chiral limit. Therefore, its representation would be only valid at $\kappa = 0$ limit. }
\end{table}

0. \emph{Preliminary} Before going into detail, let us elaborate on the symmetry representation of a given basis. For simplicity, we ignore spin for now, which reduces the number of flavors from 8 to 4. In \cite{BultnickKhalaf2020}, the sublattice basis was introduced, but here we choose the band basis. In this discussion, we focus on the following symmetries: (1) anti-unitary particle-hole symmetry (sublattice symmetry) $\cC_\textrm{sub}$, which exists only at $\kappa = 0$, (2) anti-unitary particle-hole symmetry $\cC$ from small-angle approximation, whose definition is given in \eqnref{eq:PHS}, (3) time-reversal $\cT$, and (4) composition of $C_2$ rotation and time-reversal $C_2 \cT$.
It is important to emphasize that here we examine bloch wavefunctions, and symmetry operators should be discussed in the first-quantized language. As a result, $\cC$ and $\cC_\textrm{sub}$ are \emph{unitary} operators in this context.

For a given representation of symmetries in the band basis, the microscopic action (\emph{right action}) of a unitary symmetry operator ${\cal O}$ becomes the basis action (\emph{left action}) as the following:
 \begin{equation}
     {\cal O} u_i(\bk) = {\cal O} \mqty( C^i_\bk \\ C^i_{\bk+\bG_1} \\ C^i_{\bk+\bG_2} \\ \vdots ) = u_j(\bk) \bar{{\cal O}}_{ji}
 \end{equation}
where $u_i(\bk)$ is $i$-th eigenvector at momentum $\bk$. With this understanding, the form factor $\Lambda$ and operator-inserted form factor $\Lambda_i$ should satisfy 
 \begin{align} \label{eq:sym_constraint}
     \Lambda(\bk,\bq) &= U^\dagger (\bk) U(\bk+\bq) =    U^\dagger (\bk) \qty[ {\cal O}^{-1} {\cal O}]  U(\bk+\bq) = \bar{{\cal O}}^\dagger \Lambda(\bk,\bq) \bar{{\cal O}} \nonumber \\
          \Lambda_i(\bk,\bq) &= U^\dagger (\bk) M_i U(\bk+\bq) =    U^\dagger (\bk) \qty[ {\cal O}^{-1} M_j {\cal O}]  U(\bk+\bq) = \bar{{\cal O}}^\dagger \Lambda_j(\bk,\bq) \bar{{\cal O}}
 \end{align}
where $U(\bk)$ is now a horizontally stacked eigenvectors, i.e., $N \times 4$ matrix, where $N$ is the number of microscopic degrees of freedom (layer, valley, sublattice, and moir\'e reciprocal lattice vectors) and there are valley and sublattice degrees of freedom. Here we assume that $M_i = {\cal O}^{-1} M_j {\cal O}$. Now, assume ${\cal O}$ is an anti-unitary operator, given as ${\cal O} = {\cal U} \cK$, where ${\cal U}$ is the unitary part. Assume that this anti-unitary operator has a unitary representation on basis, i.e., ${\cal O} u_i(\bk) =  u_j(\bk) \bar{{\cal O}}_{ji}$. Since $\cal O$ is anti-unitary, ${\cal O} \qty[ \alpha u_i(\bk) ]= \alpha^* u_j(\bk) \bar{{\cal O}}_{ji}$. Then,
 \begin{align} \label{eq:sym_constraint2}
      \Lambda^*(\bk,\bq) &= U^T (\bk) U^*(\bk+\bq) =    U^T (\bk) \cdot {\cal U}^\dagger {\cal U} \cdot U^*(\bk+\bq)  \nonumber \\
     &= \qty[ \,{\cal U} \cK U (\bk)]^\dagger   \qty[\, {\cal U} \cK U(\bk+\bq) ] = \bar{{\cal O}}^\dagger \Lambda(\bk,\bq) \bar{{\cal O}} 
 \end{align}
Under this band basis, we can choose the gauge such that these have the following representations: ($i$) $\cT U(\bk) = U(-\bk) \tau_x n_z$,  ($ii$) $\cC U(\bk) = U(\bk) n_x \tau_x$, and ($iii$) $(C_2 \cT) U(\bk) = U(\bk) n_z$. Two conditions that have to be enforced for the band basis are the followings: (i) $[n_z,H] = 0$ since $H$ is diagonal in the band basis, and (ii) $n_z$ anti-commutes with particle-hole symmetries $\cC$ and $\cC_\textrm{sub}$. 
To be a valid symmetry representation, symmetries should satisfy the following commutation relations originated from the microscopic representation: $\cC {\cal T} = - {\cal T} \cC$, $ \cC (C_2 {\cal T}) = -(C_2 {\cal T})  \cC$ and ${\cal T} (C_2 {\cal T}) = (C_2 {\cal T}) {\cal T}$. In \tabref{tab:sym_rep}, we listed a few different representations of symmetry operators which satisfy this set of commutation relations. In this paper, we exclusively use the band basis I in \tabref{tab:sym_rep}, but note that this is not unique; for comparison, we show another band basis representation in \tabref{tab:sym_rep}.

In the following, let us fix the basis by choosing the representation in Band Basis I. Using these symmetries, one can deduce the matrix structure of the projected Hamiltonian and the form factors. The Hamiltonian is diagonal in valley such that only terms of the form $\tau_{0,z} n_{0,x,y,z}$ are allowed. Then, $\qty{\cC,H}=0$ restricts them further to $\tau_0 n_{y,z}$ and $\tau_z n_{0,x}$. Out of these four terms, $C_2 \cT$ forbids $\tau_z n_x$. Therefore, we get
\begin{equation} \label{eq:H_chiral}
    h(\bk) =  h_0(\bk) \tau_z  + h_y(\bk) n_y + h_z(\bk) n_z 
\end{equation}
Obtaining form factor is a bit different. In this case, we use \eqnref{eq:sym_constraint} and \eqnref{eq:sym_constraint2}. Also, the coefficients in front of the Pauli matrices can be complex. We get
\begin{equation} \label{eq:Lambda_chiral}
    \Lambda_0(\bk,\bq) \sim C_0 + C_1 n_x + C_2 \tau_z n_y + C_3 \tau_z n_z
\end{equation}
where $C_0, C_2, C_3$ are real while $C_1$ is imaginary. In the chiral limit, $\kappa = 0$, the sublattice chiral symmetry $\cC_\textrm{sub} = \tau_z n_x$  further constrains $h(\bk)$ and $\Lambda_0(\bk,\bq)$, to $h_0(\bk) = 0$ and $C_2 = C_3 = 0$.
Now that we know the structure of the full interacting Hamiltonian depending on the presence of chiral symmetry $\cC_\textrm{sub}$, we can understand the global symmetry of the system as follows.

\bigskip

1. \emph{Chiral limit} ($\kappa=0$): $\hat{H}_\textrm{kin}=0$ and $\hat{H}_\textrm{int} \neq 0$: In this limit, as $\Lambda_\bq$ is given as the linear combination of $\mathbb{I}$ and $n_x$, generators of $\U(8)$ that commute with $n_x$ would still be valid symmetry generators, which halves the number of valid generators from 64 to 32. Those 32 generators can be divided into two sets of 16, each of which generate a $\U(4)$ symmetry. Therefore, the system has $\U(4)_+ \times \U(4)_-$ symmetry, where the subscript $\pm$ corresponds to the eigenvalue of $n_x$. In the sublattice basis \cite{BultnickKhalaf2020}, the operator $n_x$ corresponds to $\tau_z \sigma_z$ which is the Chern number.

\bigskip 

2. \emph{Chiral limit} ($\kappa=0$): $\hat{H}_\textrm{kin}\neq 0$ and $\hat{H}_\textrm{int} \neq 0$: In this case, a non-zero dispersion is given by $h(\bk) = h_y(\bk) n_y + h_z(\bk) n_z$. Therefore, symmetry generators must commute with $n_y$ in addition to $n_x$. This will again halve the number of symmetry generators from 32 to 16. As a result, the system has a single $\U(4)_\textrm{chiral}$ symmetry with 16 generators, which basically acts trivially on the band index and non-trivially on the spin-valley indices. 

\bigskip 

3. \emph{Non-Chiral limit} ($\kappa\neq 0$): $\hat{H}_\textrm{kin} = 0$ and $\hat{H}_\textrm{int} \neq 0$: In this case, the form factor take the following general form of \eqnref{eq:Lambda_chiral}. 
As a result, the interaction Hamiltonian is invariant only under generators that commute with $n_x$ and $\tau_z n_y$, which decreases the number of valid generators from 64 to 16. Therefore, the system has a $\U(4)_\textrm{flat}$ symmetry. Note that this $\U(4)_\textrm{flat}$ is different from $\U(4)_\textrm{chiral}$ since while the generators of flat one commute with $n_y$, those of the chiral one commute with $\tau_z n_y$.

\bigskip 

4. \emph{Non-Chiral limit} ($\kappa\neq 0$): $\hat{H}_\textrm{kin} \neq  0$ and $\hat{H}_\textrm{int} \neq 0$: Introducing the dispersion will further halve the number of generators from 16 to 8 compared to the third scenario because generators should commute with $\tau_z n_y$ in addition to $n_y$ and $n_x$. As a result, we have a $\U(2)_\bK \times \U(2)_{\bK'}$ symmetry, consistent with $4 + 4 = 8$ generators.
\bigskip

\begin{figure}[t]
    \centering
    \includegraphics[width = 0.5 \textwidth]{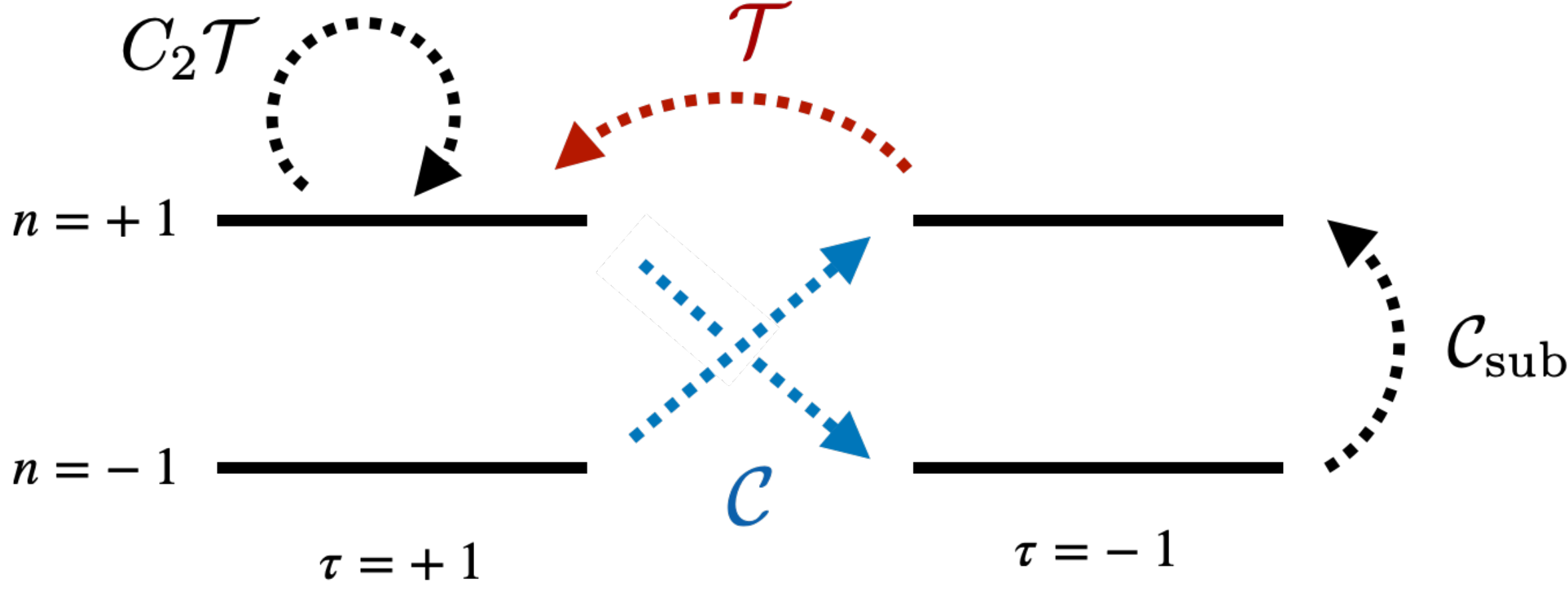}
    \caption{ Schematic diagram illustrating how symmetries in \tabref{tab:sym_rep} transforms the band basis in TBG. For the absence of sign problem, enforcing $\cC$ is most important. The diagram illustrates how we use wavefunctions at $(\tau,n) = (1,-1)$ and both $\cC$ and $\cT$ symmetries to generate the wavefunctions of all other valleys and bands in the projected subspace. However, note that conduction and valence bands are in fact degenerate at Dirac points, which is simplified in the diagram. Due to the degeneracy issue, one has to be careful to resolve band indices at degenerate $\bk$-points.   \label{fig:sym_rep}   }
\end{figure} 

\subsection{Gauge Fixing in the Simulation}

To avoid a sign-problem in the DQMC simulation, we employ the small-angle approximation which enforces the anti-unitary particle-hole $\cC$ symmetry to be exact. Furthermore, in order to establish a fair comparison among different correlation functions, it is desirable to have a well-chosen basis where the symmetry operators have a straight-forward representation, as in \tabref{tab:sym_rep}. To achieve this, we first solve the Bistritzer-Macdonald Hamiltonian in \eqnref{eq:tbg_ham}. However, the computed eigenstates have an arbitrary gauge choice at each momentum $\bk$. Hence, we need to fix the gauge as discussed in the following: 
\begin{enumerate}
    \item Choose the eigenstates for $n=\pm 1$ and $\tau=+1$ in half of the Brillouin zone, including $\bk=0$ ($\Gamma$-point). 
    \item Choose the representation of $C_2 \cT$. Then, using the microscopic action of $C_2 \cT$ symmetry, we redefine the basis at each $\bk$ from the computed eigenstates to be consistent with the representation. Assume that $M_{ij}(\bk) = \bra{\Psi_i(\bk)} {{\cal O}}_{C_2 \cT} \ket{\Psi_j(\bk)}$. Then, under the gauge transformation $\ket{\Psi_i(\bk)} \mapsto e^{i \theta_i(\bk)} \ket{\Psi_j(\bk)}$, $M_{ij}(\bk) \mapsto e^{-i(\theta_i(\bk) + \theta_j(\bk))} M_{ij}(\bk)$. For example, if we choose a band basis representation, $M_{ij}$ is diagonal, which implies that $M_{ij}$ is invariant under the gauge transformation $\ket{\Psi_i(\bk)} \mapsto \eta_i(\bk) \ket{\Psi_j(\bk)}$ with $\eta_1(\bk) = \pm 1$ and $\eta_2(\bk) = \pm 1$. If we choose a sublattice representation, then diagonal terms are vanishing while off-diagonal terms survive. In this case, any gauge transformation that satisfies $\theta_1(\bk) + \theta_2(\bk) \equiv 0$ mod $2\pi$ will leave $M_{ij}$ invariant.  Therefore, $C_2 \cT$ alone does not completely fix the gauge choice. 
    \item Choose the representation of $\cC$. Using the microscopic representation of $\cC$ that exchanges valleys but keeps $\bk$ unchanged, we can reconstruct the basis at (-)-valley for the half the Brillouin zone. While doing so, we should choose a consistent representation for $\cC$.
    \item Choose the representation of $\cT$. Using microscopic representation of $\cT$, we can construct the basis at the other half of the Brillouin zone for both $(+)$ and $(-)$ valleys. 
    \item {\bf Caveat} 
    It is important to emphasize that there can be some ambiguity at Dirac point due to the energy degeneracy. If we naively apply $C_2 \cT$ for one of the computed degenerate basis at Dirac points, it would not map it back to itself in general. Therefore, it is important to find a basis first that properly realize the representation of $C_2 \cT$ at Dirac points. 
\end{enumerate}
This will guarantee  the validity of the representations in \tabref{tab:sym_rep}, or more importantly, the structure of form factors and Hamiltonians in \eqnref{eq:Lambda_chiral}. Note that an explicit presence of chiral $\cC$ symmetry assumes that for each $\bk$-point in one valley, there should be $\bk$-point in the other valley as well. As the time-reversal symmetry assumes that for each $\bk$-point in one valley, there is $(-\bk)$-point in the other valley, the presence of both symmetries implies that our momentum grid should be symmetric with respect to the $\Gamma$-point ($\bk=0$).

\subsection{Correlation function}

In the fermionic DQMC, all order parameters and correlation functions are obtained in terms of fermionic Green's function. Let labels $i$ and $j$ stand for $i \equiv (\bk_1, \tau_1, f_1)$ and $j \equiv (\bk_2, \tau_2, f_2)$, where $f$ is electron flavor label. Note that for a given auxiliary field configuration $\{ \sigma_{r,\tau} \}$, the equal-time electron Green's function at time $\tau_1 = \tau_2 =  n \delta$ is given as the following ($0 \leq n < N_\tau$):
\begin{align}
    \expval{c^{\,}_i c_j^\dagger}_\sigma &= \expval{c^{\,}_{f_1} (\bk_1, \tau) c_{f_2}^\dagger (\bk_2, \tau) }_\sigma = \qty[ \frac{1}{1 + B_{n_2} \cdots B_{0} B_{N_\tau-1} \cdots B_{n_2+1}  } ]_{ij} \equiv \textrm{Gr}_{ij} \nonumber \\
    \expval{c_i^\dagger c_j^{\,} }_\sigma &= \expval{c^{\dagger}_{f_1} (\bk_1, \tau) c_{f_2}^{\,} (\bk_2, \tau) }_\sigma = \qty[  \frac{1}{1 + B_{n_1} \cdots B_{0} B_{N_\tau-1} \cdots B_{n_1+1}     } ]_{ji} \equiv \textrm{Grc}_{ij}
\end{align}
where $B_n$ is the matrix representation of $n$-th trotterized piece of the HS-decoupled Hamiltonian, which is responsible to evolve $\tau = n\delta$ into $\tau = (n+1) \delta$. 
The imaginary time is periodic in $\beta$, such that the $\tau$-index runs from 0 to $N_\tau-1$. Note that $c_i^\dagger c_j^{\,} = \delta_{ij} -  c_j^{\,} c_i^\dagger$; therefore, we get $\textrm{Gr}_{ij} = \delta_{ij} - \textrm{Grc}_{ji}$.

A typical operator is given as  $ {\cal O}_\alpha(\bq) \equiv  \sum_\bk c^\dagger_{\bk + \bq} \Lambda_\alpha(\bk, \bq) c^{\,}_\bk$ and the corresponding correlation function is readily calculated using Wick's theorem for a fixed configuration $\{\eta\}$:
\begin{align}
    S_\alpha(\bq,\tau) \Big|_{\{\eta\}} &= \frac{1}{\textrm{Vol}} \expval{ \qty[\sum_\bk c^\dagger(\bk+\bq, \tau) \Lambda_\alpha(\bk, \bq) c^{\,} (\bk, \tau) - \expval{O} ] \qty[\sum_{\bk'} c^\dagger(\bk'-\bq, \tau) \Lambda_\alpha(\bk', -\bq) c^{\,}(\bk', \tau) - \expval{O} ] }_{\{\eta\}}  \nonumber \\
    &= \frac{1}{\textrm{Vol}} \expval{ \qty[\sum_\bk c^\dagger(\bk+\bq, \tau) \Lambda_\alpha(\bk, \bq) c^{\,} (\bk, \tau) ] \qty[\sum_{\bk'} c^\dagger(\bk'-\bq, \tau) \Lambda_\alpha(\bk', -\bq) c^{\,}(\bk', \tau)  ] }_{\{\eta\}} - \frac{1}{\textrm{Vol}} \qty(\sum_{\{\eta\}} \expval{O}_{\{\eta\}})^2 \nonumber \\
    &= \frac{1}{\textrm{Vol}} \sum_\bk \sum_{\bk'} \sum_{ijkl} \Lambda^{ij}_\alpha(\bk, \bq) \Lambda^{kl}_\alpha(\bk', -\bq) \expval{ \qty[ c^\dagger_i(\bk+\bq, \tau)  c^{\,}_j (\bk, \tau)] } 
    \expval{ \qty[ c^\dagger_k(\bk'-\bq, \tau)  c^{\,}_l(\bk', \tau)] }_{\{\eta\}}  \nonumber \\
    &+\frac{1}{\textrm{Vol}} \sum_\bk \sum_{\bk'} \sum_{ijkl} \Lambda^{ij}_\alpha(\bk, \bq) \Lambda^{kl}_\alpha(\bk', -\bq) \expval{ \qty[ c^\dagger_i(\bk+\bq, \tau) c^{\,}_l(\bk', \tau)] } \expval{ \qty[c^{\,}_j (\bk, \tau) c_k^\dagger(\bk'-\bq, \tau)  ] }_{\{\eta\}} 
\end{align}
Note that the time-ordering is implicitly taken into account. If two order-parameter dependent form factors $\Lambda_\alpha$ and $\Lambda_\beta$ are related by the symmetry of HS-decoupled Hamiltonians, then $S_\alpha = S_\beta$ which we observe in special limits of the TBG Hamiltonian.

\section{Vanishing Quantum Hall Correlation Function in DQMC} \label{app:degeneracy}

As pointed out in \cite{BultnickKhalaf2020}, Quantum Hall phases are valid ground states at chiral limits in both flat band and dispersive cases. However, in \figref{fig:QMC_channel}, the magnitude of the quantum Hall correlation function does not exhibit the volume-law scaling behavior unlike the other observables. What is happening? 

This is the manifestation of the degeneracy of the groundstate manifold. Basically, the quantum Monte Carlo method is an importance sampling method, which explores all possible quantum states with the probability determined by the energy. At a low enough temperature, the DQMC is sampling the groundstates. As pointed out in \cite{BultnickKhalaf2020}, there are several distinct groundstate manifolds in the flat and nonflat chiral limits, distinguished by their Chern numbers \footnote{Note that the groundstates with the same Chern number can be rotated into each other by the global symmetry.} Therefore, if we measure the QH correlation function in the flat chiral limit during the QMC at $\beta \rightarrow \infty$ limit, its expectation value would be given as the following:
\begin{equation}
    \expval{\hat{S}_\textrm{QH}(0)} =  \frac{\sum_{\psi_\textrm{gs}} \expval{\psi_\textrm{gs}|\hat{S}_\textrm{QH}(0)|\psi_\textrm{gs}}}{\sum_{\psi_\textrm{gs}} 1}
\end{equation}
where the denominator counts the total groundstate degeneracy. If the majority of the groundstates have vanishing contribution to the QH channel, its correlation function may vanish in the simulation even if QH state is a valid groundstate. Following \cite{BultnickKhalaf2020}, the groundstate manifold decomposes into
\begin{equation}
    {\cal H}_\textrm{gs} =     {\cal H}_{C=0} \bigcup {\cal H}_{C=2} \bigcup  {\cal H}_{C=4}.
\end{equation}
In the continuum thermodynamic limit, each Hilbert space is parametrized by the following manifold:
\begin{align}
    {\cal H}_{C=0}  = \qty(\frac{\U(4)}{\U(2)\times \U(2)})^{\otimes 2}, \qquad 
    {\cal H}_{C=2}  = \qty(\frac{\U(4)}{\U(3)\times \U(1)})^{\otimes 2}, \qquad
    {\cal H}_{C=4}  = \mathbb{Z}_2 
\end{align}
Counting the dimensionality of the manifold, one can immediately notice that manifolds with $C \neq 0$ have  measure zero compared to the $C=0$ manifold, which implies that $\expval{\hat{S}_\textrm{QH}(0)}$ should completely vanish in thermodynamic limit. However, in  \figref{fig:QMC_channel}(a,b), we observe a finite quantum Hall correlation function in a finite size system, which exhibits an interesting scaling behavior. To understand this, we count the degeneracy of each Hilbert space for a finite size system.

As we will review below, this can be calculated using the Young-diagram approach and hook's rule \cite{PrincetonTBGIV, SUn_calculation}. To explain the surprising behavior of the QH channel, we will determine the scaling of the groundstate degeneracy with the system size. We use the band basis defined in \appref{app:C}\,2. Note that the chern number then is the eigenvalue of $n_x$, the pauli operator that flips the band index.

\subsection{Chiral limit with flat band}

In this case, we have $\U_+(4) \times \U_-(4)$ symmetry, where each $\U_\pm(4)$ acts on the subspace with different Chern number $C= n_x = \pm 1$. In each Chern sector, we have four spin-valley flavors under $\SU(4)$ representation.

First, let us consider a groundstate with net Chern number zero, i.e., the filling for each Chern sector becomes $\nu_- = \nu_+ =2$ for every momentum $\bk$. Since the electron wavefunction is anti-symmetric for the same $\bk$,  it should be antisymmetrized in the flavor index. In other words, the state with two electrons per $\bk$ should transform as the following Young diagram:  {\tiny $ \Yvcentermath1 \yng(1,1)$}. In general, tensoring two fundamental representation yields {\tiny $ \Yvcentermath1 \yng(2) \oplus \yng(1,1)$}, but the symmetric one disappears as electron creation operators anticommute. The basis of the representation {\tiny $ \Yvcentermath1 \yng(1,1)$} is represented as 
\begin{equation}
  \ket{m\equiv(\alpha,\beta)} = c^\dagger_{\bk,\alpha}c^\dagger_{\bk,\beta} \ket{0} = \frac{1}{2}\qty(c^\dagger_{\bk,\alpha}c^\dagger_{\bk,\beta} - c^\dagger_{\bk,\beta} c^\dagger_{\bk,\alpha}) \ket{0}, \qquad 1 \leq \alpha < \beta \leq 4.
\end{equation}
which is $6$-dimensional. We have the same irreducible representation (irrep.) for the other value of $\tau_z \sigma_z$. Therefore, the irrep. at $\bk$ for both Chern sectors are given by $[1^2_+, 1^2_-]$ (row length notation) which transforms under $\U(4)_+ \times \U(4)_-$.

Now, consider a particular ground-state, a Slater state represented by the single-particle density matrix $P(\bk) = (\bI + Q)/2$. In flat-band limit, any Slater state represented by $Q$ that satisfies $[Q,n_x] = 0$ is a valid groundstate. For example, we can choose $Q = \tau_z$, which corresponds to the valley-polarized (VP) state. When the system contains $N_M$ moir\'e unit cells, that electron wavefunction for the VP state belongs to the following irrep. for each $C$:
\begin{equation} \label{eq:irrep_example}
  \Yvcentermath1 \underbrace{ {\yng(2,2) \cdots\cdots \yng(2,2)} }_{N_M}
\end{equation} 
which is denoted as $[(N_M^2)_+,(N_M^2)_-]$ \cite{PrincetonTBGIV}. This is because electron creation operators at different $\bk$s are already anti-symmetrized due to the fermionic statistics, which implies that flavor labels should be symmetrized in the wavefunction. For example, if we consider the state $\prod_{\bk} (c^\dagger_{\bk,\alpha}c^\dagger_{\bk,\beta}) \ket{0}$, the flavor indices are identical across different $\bk$s, meaning that they are symmetric. Therefore, this groundstate belongs to the irrep. of \eqnref{eq:irrep_example}. 
Then, under the action of $\U(4)$ symmetry, it explores all other basis states of this irrep., which exhausts the ground state manifold. The degeneracy of the irrep. $[N_M^2]$  can be calculated by hook's formula:
\begin{equation}
  d_{[N_M^2]} = \frac{1}{12} (N_M+3)(N_M+2)^2(N_M+1) \propto N_M^4
\end{equation}
Therefore, the irrep. $[(N_M^2)_+,(N_M^2)_-] \equiv [(N_M^2)_+] \otimes [(N_M^2)_-]$ scales as $N_M^8$.

What about the groundstate manifold with $C \neq 0$? Consider the case $\nu_- = 1$ and $\nu_+ = 3$ so that there is a non-zero net Chern number of $C=2$. In this case, we see that the representation is given by $[(N_M^3)_+,(N_M)_-]$. Note that 
\begin{align}
  d_{[N_M^3]} &= \frac{1}{6} (N_M+3)(N_M+2)(N_M+1) \propto N_M^3 \nonumber \\
  d_{[N_M]} &= \frac{1}{6} (N_M+3)(N_M+2)(N_M+1) \propto N_M^3\,.
\end{align}
Hence, the total degeneracy scales as $N_M^6$. This implies that the ratio of the degeneracy of $C=0$ and $C=2$ states scales as $1/(N_M^2)$, which vanishes in thermodynamic limit. Note that $C=4$ ($C=-4$) state belongs to the trivial irrep., therefore its degeneracy is one. As a result, the quantum-Hall correlations vanish although the QH state is a proper ground state.

\subsection{Chiral limit with dispersive band}

In the chiral limit with dispersive bands, the global symmetry is given by $\U(4)_\textrm{chiral}$ which acts on spin-valley indices and leaves the Chern index $n_x$ invariant (See \appref{app:C}\,2). 
Note that the relevant Hamiltonian is defined in \eqnref{eq:H_chiral} and \eqnref{eq:Lambda_chiral}. The dispersion $h \sim h_y n_y + h_z n_z$ anti-commutes with the Chern number $n_x$, which induces a second-order energy correction originating from the hopping terms. Such a correction corresponds to an anti-ferromagnetic coupling between two different Chern sectors with identical spin-valley flavor.
In terms of single-particle density matrix, on top of satisfying $[Q,n_x] = 0$, $Q$ should additionally satisfy $\{Q, n_z\}=0$. This means that for a given spin-valley flavor, if $n_x=+1$ is filled then $n_x = -1$ is favored to be empty \cite{BultnickKhalaf2020}. For a valid groundstate, we can consider $Q = \tau_z n_x$. In this configuration, for $C=n_x =1$ sector we occupy $(s_z,\tau_z,n_x) = (\pm 1, 1, 1)$ and for $C=n_x=-1$ sector we occupy $(s_z,\tau_z, n_x) = (\pm 1, -1, -1)$. This is valley-Hall (VH) insulating phase.

Which irrep. does this state belong to? Unlike the flat non-chiral case, here at each $\bk$ four electrons occupy all different spin-valley flavors, and anti-ferromagnetic interaction prefers certain superposition among them. Although it is clear that two electrons in the $C=1$ sectors and two electrons in $C=-1$ sectors are anti-symmetrized respectively, it is unclear how the combined state symmetrize or antisymmetrize the flavors.

Naively, anti-ferromagnetic interaction implies that one may relabel $C=-1$ spin-valley indices so that it becomes equivalent to $C=1$ spin-valley indices. This is reminiscent of defining Neel order as a staggered order parameter. Indeed, the presence of antiferromagnetic interaction naturally induces an anti-correlation between spin-valley occupations between $C=1$ and $C=-1$ sectors. Since the occupation of $(s_z,\tau_z)$ for $C=1$ is determined by the occupation of $(s_z,\tau_z)$ for $C=-1$, one may think that we can simply count the degeneracy for the irrep. for $C=1$ sector, defined by $[N_M^2]$, as the $C=-1$ state is completely determined by it. However, this is misleading. If we think about the familiar Heisenberg anti-ferromagnet, it maximizes the sublattice polarization $\abs{S^\textrm{tot}_A}$ and $\abs{S^\textrm{tot}_B}$. However, sublattice polarization for $A$ and $B$ forms a singlet in the ground-state, resulting in a unique ground state. This is because the anti-ferromagnetic groundstate minimizes the quadratic casimir operator $(S_A + S_B)^2$. For example, for two $\SU(2)$ spins with antiferromagnetic interaction, the groundstate is neither $\ket{\uparrow \downarrow}$ nor $\ket{\downarrow\uparrow}$, but $\ket{\uparrow \downarrow} - \ket{\downarrow\uparrow}$.

Similarly, in this case the anti-ferromagnetic interaction between different Chern-sectors imply that we want to minimizes the quadratic Casimir operator after combining the irreps. of $C=1$ and $C=-1$.  To minimize the value of the quadratic Casimir operator, it should become a trivial irrep. In other words, the ground state is represented by the following young diagram:
\begin{equation}
  \Yvcentermath1  \underbrace{ {\yng(2,2) \cdots\cdots \yng(2,2)} }_{N_M} \, \otimes \, \underbrace{ {\yng(2,2) \cdots\cdots \yng(2,2)} }_{N_M} \quad \Rightarrow \quad  \underbrace{ {\yng(2,2,2,2) \cdots\cdots \yng(2,2,2,2)}  }_{N_M} = \bm{1}
\end{equation}
where the first tableau is for $C=1$ and the second tableau is for $C=-1$.
However, as in the case of $\SU(2)$ anti-ferromagnet, there is an Anderson tower of states, which should give a huge number of states spaced by $\sim 1/V$ ($V$ is volume of the system). 
The first few excitations would have the following representations:
\begin{equation}
  \Yvcentermath1 \underbrace{ {\yng(1,1,1,1) \cdots\cdots \yng(4,3,3,2)}  \hspace{-1em} }_{N_M} \hspace{1em} = \bm{15}, \qquad     \underbrace{ {\yng(1,1,1,1) \cdots\cdots \yng(4,4,2,2)}  \hspace{-1em} }_{N_M} \hspace{1em} = \bm{20},
\end{equation}
Note that when we combine $[N_M^2]$ and $[N_M^2]$ irreps., there are some important rules. Therefore, the general irrep. of the combination is given as the following shape:
\begin{equation}
  \centering
  \scalebox{.85}{
  \begin{tabular}{r@{}l}
  &
  \begin{ytableau}
  ~               & ~ & \none[\cdots] & ~ & ~& \none[\cdots] & ~ & ~ &\none[\cdots] & ~ & ~ &\none[\cdots] & ~\\
  ~               & ~ &\none[\cdots] & ~ & ~ & \none[\cdots] & ~ & ~ &\none[\cdots] & ~ \\
  ~               & ~ &\none[\cdots] & ~ & ~ & \none[\cdots] & ~     & \none \\
  ~               & ~ &\none[\cdots] & ~ & \none  & \none  & \none      & \none 
  \end{ytableau}\\[-1.5ex]
  &$\underbrace{\hspace{6.3em}}_{\displaystyle N_M - n}$ $\underbrace{\hspace{4.2em}}_{\displaystyle \vphantom{N_M} n-m}$ $\underbrace{\hspace{4.2em}}_{\displaystyle \vphantom{N_M} 2m}$ $\underbrace{\hspace{4.2em}}_{\displaystyle \vphantom{N_M} n-m}$
  \end{tabular}
  }   = (n-m,2m,n-m )
\end{equation}
where $m \leq n$.

What about for $C=2$ groundstate? In this case, we can occupy $C=1$ sector with three spin-valley indices and $C=-1$ sector with the other spin-valley index to minimize the antiferromagnetic energy. Again, if we want to minimize the quadratic casimir, there is only the following choice:
\begin{equation}
  \Yvcentermath1  \underbrace{ {\yng(2,2,2) \cdots\cdots \yng(2,2,2)} }_{N_M} \, \otimes \,  \underbrace{ {\yng(2) \cdots\cdots \yng(2)} }_{N_M} \quad \Rightarrow \quad  \underbrace{ {\yng(2,2,2,2) \cdots\cdots \yng(2,2,2,2)} }_{N_M} = \bm{1}
\end{equation}
where the first tableau is for $C=1$ ($\nu_+ = 3$) sector and the second tableau is for $C=-1$ ($\nu_-=-1$) sector. However, there will be a tower of states as well, which will be obtained by moving each box in the last row to the arbitrary position on the right side, one by one from right to left. For example, a first few Anderson tower excitations are the following:
\begin{equation}
  \Yvcentermath1 \underbrace{ {\yng(1,1,1,1) \cdots\cdots \yng(4,3,3,2)}  \hspace{-1em} }_{N_M} \hspace{1em} = \bm{15}
  ,  \qquad   \Yvcentermath1 \underbrace{ {\yng(1,1,1,1) \cdots\cdots \yng(5,3,3,1)}  \hspace{-2em} }_{N_M} \hspace{2em} = \bm{84}
\end{equation}
Note that combining the $[N_M^3]$ and $[N_M]$ irreps., cannot create a new column of length greater or equal to $2$ since then antisymmetrizing symmetric indices is impossible \footnote{\url{http://cftp.ist.utl.pt/~gernot.eichmann/2014-hadron-physics/hadron-app-1.pdf}}. Therefore, the general irrep. of the combination is given by the following shape:
\begin{equation} 
  \centering
  \scalebox{.85}{
  \begin{tabular}{r@{}l}
  &
  \begin{ytableau}
  ~               & ~ & \none[\cdots] & ~ & ~& \none[\cdots] & ~ & ~ &\none[\cdots] & ~ \\
  ~               & ~ &\none[\cdots] & ~ & ~ & \none[\cdots] & ~     & \none \\
  ~               & ~ &\none[\cdots] & ~ & ~ & \none[\cdots] & ~     & \none \\
  ~               & ~ &\none[\cdots] & ~ & \none  & \none  & \none      & \none 
  \end{ytableau}\\[-1.5ex]
  &$\underbrace{\hspace{6.3em}}_{\displaystyle N_M - n}$ $\underbrace{\hspace{4.2em}}_{\displaystyle \vphantom{N_M} n}$ $\underbrace{\hspace{4.2em}}_{\displaystyle \vphantom{N_M} n}$
  \end{tabular}
  }  = (n,0,n)
\end{equation}

Finally, the Anderson-tower energy is approximated by 
\begin{equation}
  \delta E \sim \frac{J}{N_M} \cdot C_2, \qquad C_2: \textrm{ quadratic Casimir operator}
\end{equation}
For the Young diagram defined by $(k_1,k_2,k_3)$, the $C_2$ is calculated by:
\begin{equation}
  C_{2,\textrm{SU(4)}}(k_1,k_2,k_3) = \frac{1}{8}(3 k_1^2 + 4 k_2^2 + 3 k_3^2 + 4 k_1 k_2 + 2 k_1 k_3 + 4 k_2 k_3 + 12 k_1 + 16 k_2 + 12 k_3)
\end{equation}
and the dimension is given as 
\begin{equation}
  \textrm{dim}_{\textrm{SU(4)}}(k_1,k_2,k_3) = \frac{1}{12}\qty(k_1 +1)(k_2+1) (k_3+1) (k_1+k_2+2)(k_2 + k_3 + 2)(k_1+k_2+k_3+3)
\end{equation}
In general, if we set the threshold energy $\delta E$ as some constant and count the total dimensions of the irreps. within the energy window $[0,\delta E]$, in the $C_\textrm{tot} = 0$ case the number of Anderson tower states scales as $N_M^4$ while in the $C_\textrm{tot} = 2$ case it scales as $N_M^3$. (This can be shown in the asymptotic expansion or numerically. If we think about $\delta E$ as some temperature scale, then this indeed gives us the degeneracy that can be observed in the finite temperature simulation. Therefore, we obtain the scaling of the ratio $1/N_M$, which matches with our observation.

\section{Details of Quantum Monte Carlo, Trotter Decomposition and Greens function data} 
\label{app:QMCdetails}

\begin{figure}[t]
    \centering
    \includegraphics[width = 0.98 \textwidth]{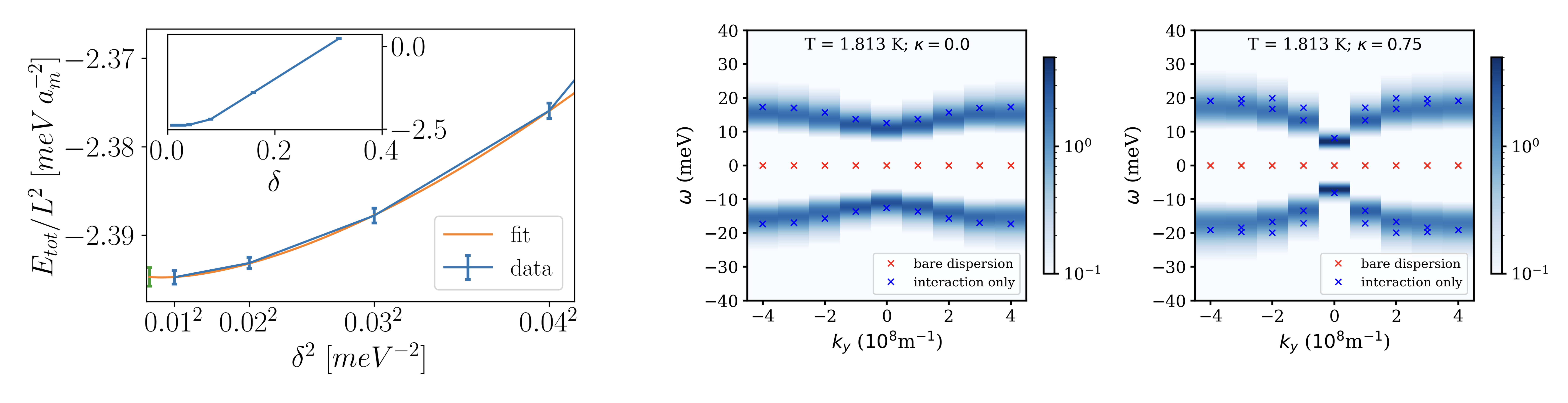}
    \caption{\emph{Left:} Convergence of the energy density with Trotter step size $\delta$. Here, we used the lowest temperature of $T=1.8\TK$ and $\kappa=0.3$ for the $L=6$ system. \emph{Right:} Comparison the the single-particle spectra obtained via MaxEnt from the DQMC simulation with exact results in the flat-band limit for (a) $\kappa=0.0$  and (b) $\kappa=0.75$  at $T=1.8\TK$. The orange and green dots represent the non-interacting flat-band limit and the exact excitation energy of the interaction, respectively. The good agreement demonstrates the correctness of our DQMC implementation.   \label{fig:flat}   }
\end{figure} 

It was necessary to overcome a few challenges to implement an efficient DQMC algorithm and we present some of the tricks that are used below.

Note that the Trotter decomposition presented in \eqnref{eq:Z} of the main text is non-hermitian such that the Trotter error can give rise to nonphysical results, e.g., negative compressibility. This is well-known and usually cured by a symmetric decomposition. Here, we go one step further and divide all interaction terms, $\hat{V}_m$, into two different groups, $\mathcal{V}_<$ and $\mathcal{V}_>$, that contain the momenta with $|\bq_m|\leq q_{\mathrm{max}}$ and $|\bq_m|>q_{\mathrm{max}}$, respectively. This separation is motivated by the momentum dependence of the Coulomb potential $V_\bq = V_0 \tanh(\abs{\bq} d)/\abs{\bq}$ with $V_0 = \frac{e^2}{2 \epsilon \epsilon_0}\frac{1}{A}$ and the form factor $\Lambda(\bk,\bq)$ 
which are peaked at $|q|=0$ such that $\mathcal{V}_<$ contains the strongly interacting terms while $\mathcal{V}_>$ includes the remaining weaker ones. This allows a refined Trotter decomposition,
\begin{equation}
    e^{-\delta \hat{H}} = 
    e^{-\frac{\delta}{2} \hat{H}_0} 
    \prod_{\substack{i=1 \\ \hat{V}_i \in \mathcal{V}_>}}^{N_>} e^{-\frac{\delta}{2} \hat{V}_i} 
    \prod_{\substack{i=1 \\ \hat{V}_i \in \mathcal{V}_<}}^{N_<} e^{-\frac{\delta}{4} \hat{V}_i} 
    \prod_{\substack{i=N_< \\ \hat{V}_i \in \mathcal{V}_<}}^{1} e^{-\frac{\delta}{4} \hat{V}_i} 
    \prod_{\substack{i=1 \\ \hat{V}_i \in \mathcal{V}_<}}^{N_<} e^{-\frac{\delta}{4} \hat{V}_i} 
    \prod_{\substack{i=N_< \\ \hat{V}_i \in \mathcal{V}_<}}^{1} e^{-\frac{\delta}{4} \hat{V}_i}
    \prod_{\substack{i=N_> \\ \hat{V}_i \in \mathcal{V}_>}}^{1} e^{-\frac{\delta}{2} \hat{V}_i} 
    \,e^{-\frac{\delta}{2} \hat{H}_0} + \mathcal{O}(N^\alpha \delta^\beta)\, ,
\end{equation}
where the weak interaction terms of $\mathcal{V}_>$ appear twice as usual in a symmetric decomposition, while the strong interactions of $\mathcal{V}_<$ appear four times, in reverted order. This decomposition scheme has multiple advantages. First, the product is hermitian as in any conventional symmetric decomposition. Second, the effective Trotter step size of the strong interaction terms is $\frac{\delta}{4}$ and thus the related Trotter error is considerably smaller which is particularly helpful here due to the unfavourable scaling with system size $N$. The convergence of the Trotter error is shown in \figref{fig:flat}. Third, we introduce more auxiliary fields for the strong coupling terms that are therefore updated more often such that we spend the computational effort where it matters. Finally, the term with $\bq=0$ now appears four times on each time slice. This allows use to choose both continuous field for the inner two appearances as well as discrete auxiliary fields for the outer ones. This solves sampling issues, namely long auto-correlation and warm-up times, due to non-efficient updates where the flipping the discrete field may change the configuration too much and is only acccepted very rarely or small updates of the continuous field is accepted often, but doesn't really change the configuration. Having both updates available allows the stochastic process at the heart of the DQMC algorithm to choose between the two schemes as needed. This is particularly important when the non-interaction band is (perfectly) flat. The discrete HS fields $\eta$ may take four distinct values as listed below along with the according weight $f$ \cite{Motome97,Assaad97},
\begin{equation} \label{eq:eta_gamma_fields}
\begin{aligned}
\eta &= \pm \sqrt{2 \left(3 - \sqrt{6} \right)}, \quad & f[\eta] &= 1 + \sqrt{6}/3,\\
\eta &= \pm \sqrt{2 \left(3 + \sqrt{6} \right)}, \quad & f[\eta] &= 1 - \sqrt{6}/3.
\end{aligned}
\end{equation}

From a physics perspective, the $V_{\bq=0}$ contribution of the Coulomb repulsion couples to the total charge, $(\bar{\rho}_{\bq=0}-\rho^0)^2$, measured with respect to charge neutrality, $\rho^0=4N_M$. Since the total particle number is a conserved quantity, this contribution only raises the energy of particle sectors different from half filling. For example, adding an additional electron, as probed by the single-particle greens function $G(\tau)$ and the corresponding spectra $A(\omega)$, costs a finite amount of energy, $V(\bq=0)$, which vanishes in the thermodynamic limit due to the $\frac{1}{\mathrm{vol}}$ normalization in $V_0$. However, this term may introduce a large charging energy on small lattices, in particular for large gate distances, $d_s$.
To improve the convergence to the thermodynamic limit, we averaged the Coulomb potential over a hexagonal plaquette that are constructed such that the hexagons of neighboring momenta touch, $V_\bq \mapsto \frac{1}{A_{\hexagon}} \sum_{\tilde{\bq} \in \hexagon_\bq} V_{\tilde{\bq}} $. Note that this improved weight of the interaction converges to the original value of $V_\bq$ for large lattices, while improving the finite size convergence of both the charging energy and the bare dispersion band width.

\begin{figure}
    \centering
    \includegraphics[width = 0.6 \textwidth]{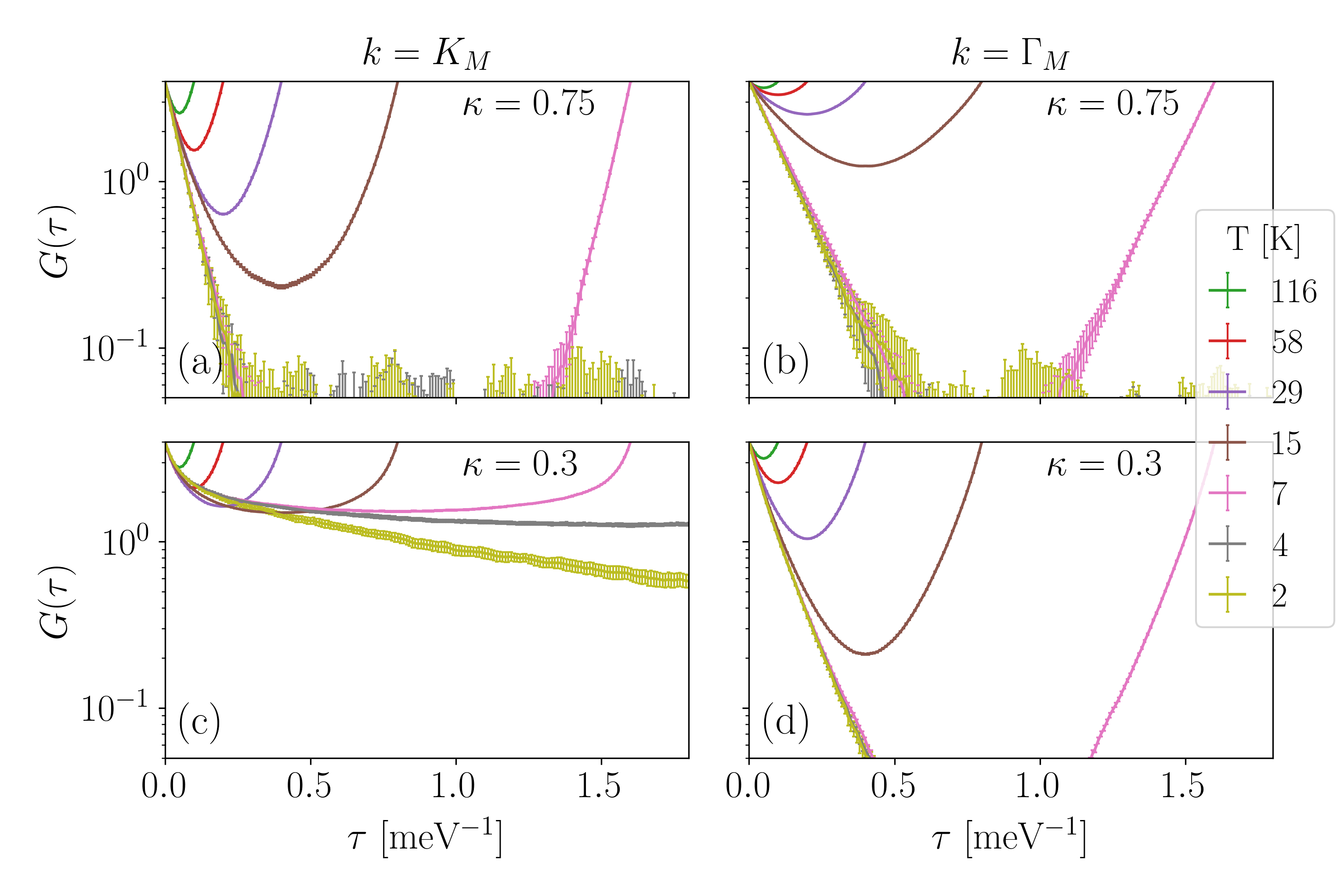}
    \caption{ Greens function data used by the maximum entropy method to extract the spectra shown in \figref{fig:SpecCut} of the main text.  \label{fig:Gtau}   }
\end{figure} 

Additionally, we average over the shifted Brillouin zones for the single-particle momentum $\bk$ that are compatible with the symmetries, i.e., contain both $\bk$ and $-\bk$. This is equivalent to inserting a $\pi$-flux through the holes of the torus in a conventional real-space lattice setup and is also known as averaging over periodic and anti-periodic boundary conditions. This improves the finite size convergence of quantities that are sensitive to the single-particle momentum grid, such as the density of states or the compressibility. Observables, whose finite size effects are dominated by finite size gaps of two-particle operators, e.g., the correlation length, are less improved since the grid for $\bq=\bk-\bk'$ is unchanged by the flux insertion.

We would like to make a very technical remark on the memory consumption of the algorithm before closing this section with a final sanity check on the method. The interaction, formulated in momentum space, involves the degrees of freedom at every momentum $\bk \in \mathrm{BZ}$, i.e., $\bar{\rho}_{\bq} = \sum_{\bk, \bk' \in \mathrm{BZ}} c^\dagger_{\bk'} H^{\bq}_{\bk',\bk} c^{\,}_{\bk} - \rho^0$ where the matrix $H^{\bq}_{\bk',\bk}$ and the potential background shift $\rho^0$ can be inferred from \eqnref{eq:denOps} of the main text. While these matrices are sparse, $e^{H^{\bq}_{\bk',\bk}}$ is a dense $2L^2 \times 2L^2$-matrix\footnote{We employed the block-diagonal structure in both valley and spin. Otherwise, the linear dimension would increase by a factor of up to 4.}. This the exponentiation is numerically expensive, it is beneficial to calculate them once at the beginning of the simulation and store them such that they are readily available during the remainder of the simulation. However, the unusually large dimension requires a comparatively huge amount of RAM\footnote{For the Hubbard interaction, the matrix actually is only a scalar number or a $2\times 2$ matrix at the most.}. We were able to overcame this issue to using shared memory between different MPI-tasks on the same compute node of the cluster.

To close this section, we compare the single-particle spectra for both the chiral limit, $\kappa=0$, and away from it, $\kappa=0.75$, for the flat band limit. Since the interaction is diagonal in valley space, it is straight forward to identify the fully valley polarized state as an exact eigenstate of the Coulomb interaction with vanishing energy as listed in \tabref{tab:OP}. It is also straight forward to identify the addition of a single electron as an exact eigenstate, giving rise to the purely interacting dispersion depicted in \figref{fig:flat} along with the spectral function extracted via MaxEnt from the DQMC simulation. The good agreement of the results confirms that the algorithm does not only reproduce the ground state hierarchy, see sec.~\ref{sec:GShierarchy}, but also finds the correct excitation energies for the flat-band limit, an thus proves the correctness of our implementation.

Extracting spectral function via the maximum entropy method is subtle due to the ill-defined underlying analytic continuation. However, the spectral features that we presented in the main text, \figref{fig:sum} and \figref{fig:SpecCut}, are also clearly visible in the raw data for $G(\tau)$ as shown in \figref{fig:Gtau}. Typically, $G(\tau)$ decays exponentially for imaginary times $\tau$ close to $\tau=0$ or $\tau=\beta/2$. Since the spectrum is particle-hole symmetric $A(\omega)=A(-\omega)$, the single-particle Greens function satisfies $G(\tau)=\int_0^\infty d\omega \frac{1}{\pi}\frac{\cosh((\tau-\beta/2)\omega)}{\cosh(\beta/2\omega)} A(\omega)$, and the data of \figref{fig:Gtau}(a,b,d) strongly suggests a (thermally broadend) peak at finite frequency $\omega=\pm \varepsilon$. The data in panel (c) is different and shows a fast decaying mode at short times, $\tau <0.2\meV^{-1}$, indicating a high energy mode, as well as a low energy mode with an energy set by the slow decay rate for $\tau>0.2\meV^{-1}$. Note that this low energy mode can only be resolved if $\beta>0.4\meV^{-1}$. One can extrapolate the slowly decaying exponential at large times to $\tau=0$ to extract the total spectral weight of the low energy mode. The weight is estimated to be of order 2.
All of these feature are captured by the maximum entropy method as well as presented in \figref{fig:sum}, \figref{fig:SpecCut} and \figref{fig:flat}.

\begin{figure}[t]
    \centering
    \includegraphics[width = 0.4 \textwidth]{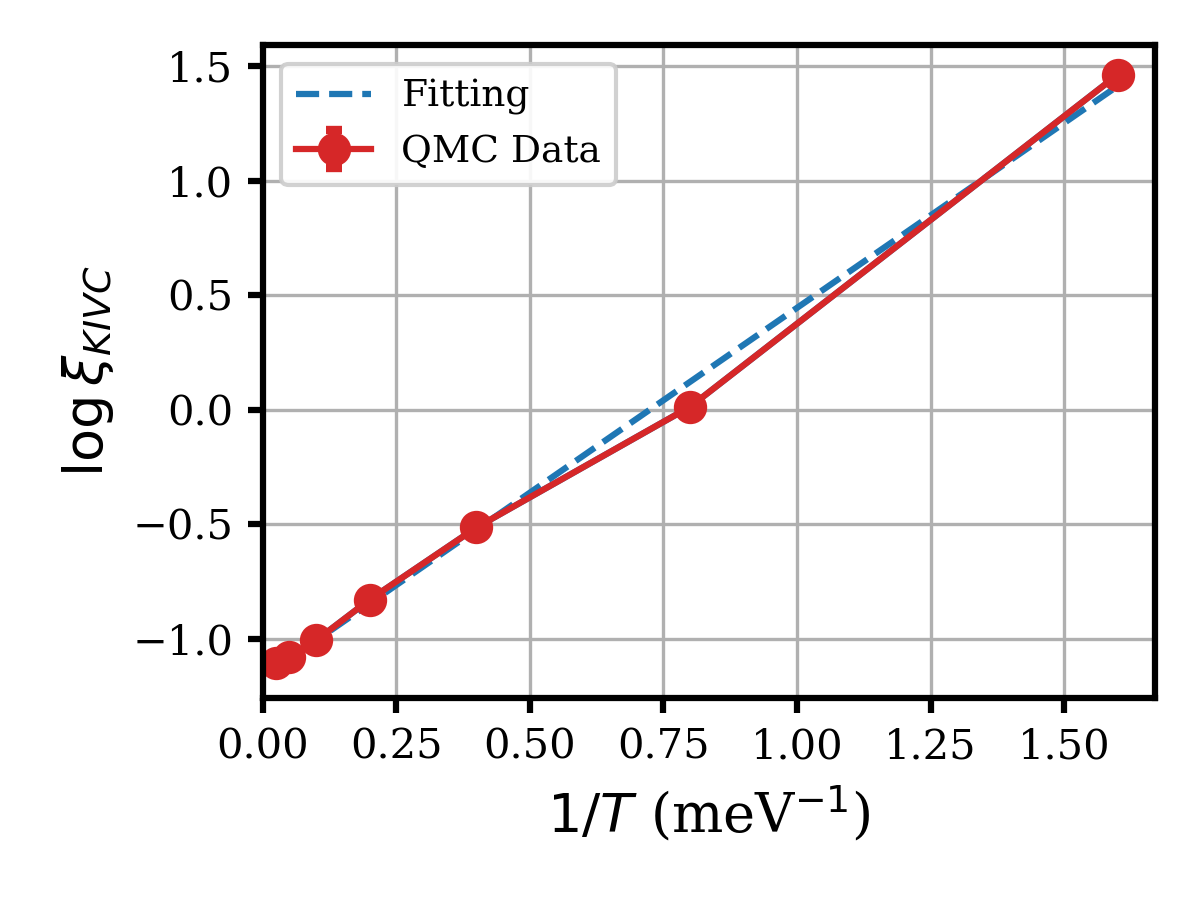}
    \caption{  \label{fig:stiffness} Here we plot $\log \xi_\textrm{KIVC}$ against the inverse of temperature $1/T$. Using the relation in \eqnref{eq:corr_stiff}, we extract the stiffness to be $\rho \sim 1.03 \meV$.  }
\end{figure} 

\section{Correlation length from the non-linear sigma model}
\label{App:correlationLength}

In this appendix, we show how the temperature dependence of the correlation length can be used to extract the stiffness of the non-linear sigma model. 

Let us first recall the derivation of the correlation length in the standard sigma model describing a ferromagnet or an antiferromagnet whose target space is a 2-sphere. Writing the sigma model as
\begin{equation}
    S = \frac{1}{2f} \int d^2 \br (\nabla \bn)^2, \qquad f = \frac{T}{\rho}
\end{equation}
We know that, due to Mermin-Wagner theory, the vector $\bn$ cannot order at finite temperature. As a result, the correlation function $\langle n_i(\br) n_j(\br') \rangle $ decays exponentially in the distance $|\br - \br'|$ with a correlation length $\xi(T)$ that only diverges at $T = 0$. The temperature dependence of $\xi(T)$ can be obtained by solving the RG equations of the coupling $f$ at two-loop level derived by Polyakov \cite{Polyakov75}
\begin{equation}
    \frac{d f}{d \ln \Lambda} = -\frac{1}{2\pi}f^2 + O(f^3)
\end{equation}
which leads to the correlation length
\begin{equation}
    \xi = \Lambda^{-1} e^{\frac{2\pi}{f}} = \Lambda^{-1} e^{\frac{2\pi \rho}{T}}
\end{equation}
where $\Lambda$ is a UV momentum cutoff that can be identified with the inverse lattice constant (up to numerical prefactors).

These results can be straightforwardly generalized to our case by using the two loop RG equations for $f$ for the different sigma model manifolds relevant to our problem. Let us start with the chiral flat band limit with the sigma model given by
\begin{equation}
    S = \frac{1}{8f} \int d^2 \br [\tr (\nabla Q_+)^2 + \tr (\nabla Q_-)^2], \qquad f = \frac{T}{\rho}
\end{equation}
Each of $Q_+$ and $Q_-$ belong to the manifold $\frac{\U(4)}{\U(2) \times \U(2)}$. The two-loop calculation for the unitary Grassmanian manifold $\frac{\U(2N)}{\U(N) \times \U(N)}$ was done in the works of Hikami \cite{Hikami81} and Wegner \cite{Wegner89} (a concise summary of the results for any symmetric space is provided in Table III of Ref.~\cite{EversMirlin}) where
\begin{equation}
    \frac{d f}{d \ln \Lambda} = -\frac{N}{2\pi} f^2 + O(f^3)
    \label{Beta}
\end{equation}
Substituting $N=2$ and taking into account the two Chern sectors gives the correlation length
\begin{equation}
    \xi = \Lambda^{-1} e^{\frac{\pi}{f}} = \Lambda^{-1} e^{\frac{\pi \rho}{T}}
    \label{Xi}
\end{equation}
The chiral non-flat and non-chiral flat limits are obtained by setting $Q_+ = -Q_-$ and $Q_+ = \tau_z Q_- \tau_z$, respectively. In both cases, the action becomes
\begin{equation}
    S = \frac{1}{4f} \int d^2 \br \tr (\nabla Q)^2, \qquad f = \frac{T}{\rho}
    \label{SQ}
\end{equation}
with $Q \in \frac{\U(4)}{\U(2) \times \U(2)}$. Using Eq.~\ref{Beta} (note here that $f$ is rescaled by a factor of 2 but there is a single sector so there is no summation over sector, thus the final expression looks exactly the same), this gives the same expression (\ref{Xi}) for the correlation length.

Finally, let us consider the non-flat non-chiral limit. In this case, the sigma model manifold is  $\U(2)$ where $Q$ can obtained from that of the chiral non-flat limit by imposing the condition $\{Q, \tau_z\} = 0$ leading to
\begin{equation}
    Q = \left(\begin{array}{cc}
    0 & U \\ U^\dagger & 0 \end{array}\right)_{K/K'}
\end{equation}
Substituting in (\ref{SQ}) yields the action
\begin{equation}
    S = \frac{1}{2f} \int d^2 \br \tr( \nabla U^\dagger \nabla U)^2, \qquad f = \frac{T}{\rho}
\end{equation}
This manifold can be decomposed as $\U(1) \times \SU(2)$ up to a discrete $\Z_2$ identification. The dynamics of the $\U(1)$ part is very different from the $\SU(2)$ since it can undergo a BKT transition. In particular, this means that we can write a separate stiffness term for the $\U(1)$ and $\SU(2)$ parts and they will have independent correlation length \footnote{This is in a sense similar to distinguishing the charge and spin order in a spin-triplet superconductor. While the former is associated with a diverging correlations function at finite $T = T_{\mathrm{BKT}}$, the correlations function of latter only diverges at $T = 0$.}. We are interested in the correlation length for the SU(2) part which only diverges at $T = 0$. This will be governed by a sigma model whose manifold is SU(2) which is isomorphic to the 3-sphere. The beta function for the manifold $S^N$ is given by \cite{Polyakov75}
    \begin{equation}
    \frac{d f}{d \ln \Lambda} = -\frac{N-1}{\pi} f^2 + O(f^3)
\end{equation}
which gives the correlation length for our case $N=3$
\begin{equation} \label{eq:corr_stiff}
    \xi = \Lambda^{-1} e^{\frac{\pi}{2f}} = \Lambda^{-1} e^{\frac{\pi \rho}{2T}}
\end{equation}

To extract the stiffness from our numerical results, we need to identify a scale $T_c$ at which the correlation length becomes comparable to the system size $\xi \sim L$ which can then be used to extract the stiffness $\rho$ according to the expressions derived above.  
\end{document}